\documentclass[12pt]{article}
\usepackage{jheppub}
\usepackage[utf8]{inputenc}
\usepackage[T1]{fontenc}
\pdfoutput = 1

\usepackage{amssymb}
\usepackage{mathrsfs}
\usepackage{amsmath}
\usepackage{fancyhdr}
\usepackage{enumitem}
\usepackage{tikz}
\usepackage{cancel}
\usepackage{graphicx,wrapfig}
\usepackage[font=small,labelfont=bf]{caption,subcaption}

\usepackage{mathabx}
\usepackage{mathtools}
\usepackage{pdfsync}
\usepackage{slashed}
\usepackage[normalem]{ulem}
\usepackage{upgreek}
\usepackage{url}
\usepackage{enumitem}

\usepackage[ragged]{footmisc}
\usepackage{color}
    \definecolor{darkgreen}{rgb}{0,0.5,0}
    \definecolor{darkred}{rgb}{0.5,0,0}
    \definecolor{darkblue}{rgb}{0,0,0.6}
    \definecolor{purple}{rgb}{0.4,.2,0.7}

\newcommand{\bigO}[1]{\ensuremath{\mathcal{O}\left(#1\right)}}
\newcommand{\inv}[1]{\frac{1}{#1}}
\newcommand{\abs}[1]{\lvert #1 \rvert}
\newcommand*\diff{\mathop{}\!\mathrm{d}}
\newcommand{\Tr}{\text{Tr}\,}
\newcommand\der[2][]{\ensuremath{\frac{\diff#1}{\diff#2}}}

\numberwithin{equation}{section}

\begin{document}

\title{Black Tunnels and Hammocks}
\author{William~D.~Biggs,}
\author{Jorge~E.~Santos}

\affiliation{Department of Applied Mathematics and Theoretical Physics, University of Cambridge,\\
Wilberforce Road, Cambridge, CB3 0WA, UK}

\emailAdd{wb304@cam.ac.uk}
\emailAdd{jss55@cam.ac.uk}

\abstract{
We construct the holographic duals to a large $N$, strongly coupled $\mathcal{N}=4$ super Yang-Mills conformal field theory defined on a four-dimensional de Sitter-Schwarzschild background. There are two distinct five-dimensional bulk solutions. One, named the black tunnel, is static and possesses two disconnected horizons. The other, the black hammock, contains only one horizon in the bulk. The hammock horizon is not a Killing horizon, and hence possesses interesting properties, such as non-vanishing expansion and shear, as well as allowing classical flow along it. The DeTurck method was used in order to attain the black tunnel solutions, whilst the black hammocks were found in Bondi-Sachs gauge.}

\maketitle



\section{Introduction}
One of the most fascinating results in theoretical physics was the discovery of Hawking radiation \cite{Hawking:1974rv}; that quantum fields radiate on fixed black hole backgrounds. Famously, this phenomenon leads to the information paradox \cite{Hawking:1976ra}, posing deep and fundamental questions about the way a theory of quantum gravity must behave. However, for the most part, the study of Hawking radiation and, more generally, quantum field theory in curved spacetime has been focused on free or weakly interacting fields. A natural question is to ask whether taking the quantum fields to be strongly coupled affects the way they behave on a curved background, though the two difficulties of strong coupling and curved space make it extremely challenging to make any headway tackling this problem via first-principles field theory calculations.

Fortunately, gauge/gravity duality has provided us an approach to study precisely such a system. The AdS/CFT correspondence \cite{Maldacena:1997re,Witten:1998qj,Aharony:1999ti} describes a duality between an $\mathcal{N} = 4$ super Yang-Mills (SYM) conformal field theory (CFT) on a fixed, but possibly curved, manifold $\partial \mathcal{M}$ and type IIB string theory on an asymptotically locally anti-de Sitter (AlAdS) spacetime, $\mathcal{M}$, which has conformal boundary given by $\partial \mathcal{M}$. Taking the limit where the CFT possesses a large number of colours, $N$, and is strongly coupled (specifically taking the 't Hooft coupling to infinity) corresponds to taking a classical limit on the gravitational side. Hence, in order to study how such a large $N$, strongly coupled CFT behaves on a manifold $\partial M$, one can find AlAdS solutions to the Einstein equation with conformal boundary $\partial M$. See, for example, \cite{Marolf:2013ioa} for an excellent review on this method.

Much work has been done in the case where the boundary metric $\partial M$ is taken to be the Schwarzschild metric, \cite{Hubeny:2009ru, Hubeny:2009kz,Hubeny:2009rc,Caldarelli:2011wa, Figueras:2011va,Santos:2012he, Santos:2014yja,Fischetti:2016oyo, Santos:2020kmq}, and two classes of gravitational duals have been found. One class of these solutions are called black droplets, which contain two, disconnected horizons in the bulk: a horizon which extends from the boundary black hole and a perturbed planar black hole deep within the bulk. The other class are called black funnels, and these contain a single, connected horizon in the bulk which extends from the horizon of the boundary black hole into the bulk and into an asymptotic region. 

The two solutions correspond to different phases of the CFT on the Schwarzschild background. The connected, funnel solution corresponds to a \textit{deconfined} phase. In the bulk, the single horizon allows flow along it at a classical level (it is a non-Killing horizon, evading rigidity theorems due to its non-compact nature), meaning that on the field theory side there is Hawking radiation from the horizon to the asymptotic region of order $\bigO{N^2}$. On the other hand, the disconnected, droplet solution corresponds to a \textit{confined} phase; there is no classical flow between the two bulk horizons, meaning that on the field theory side the Hawking radiation of the CFT is greatly suppressed to $\bigO{N^0}$. If one were to instead excite $N^2$ free fields on a black hole background, one would always expect $\bigO{N^2}$ radiation from the horizon, and hence, this confined phase is a novel property of strongly coupled fields. The field theory mechanism behind this behaviour is still not well understood.

In such a set up, one can dial the temperature, $T_H$, of the boundary black hole and the temperature, $T_\infty$, of the limiting thermal CFT at infinity. It is interesting to investigate which of these phases dominates across values of the dimensionless parameter $T_\infty / T_H$. In \cite{Santos:2020kmq}, evidence was provided that for small values of $T_\infty / T_H$, the droplet phase dominates, whilst for large values of $T_\infty / T_H$, the funnels phase dominates.

Moreover, in \cite{Fischetti:2012ps,Fischetti:2012vt,Marolf:2019wkz}, similar AlAdS solutions were found with AdS black holes on the boundary, and similar results were found regarding the phases of the CFT on such backgrounds.

In this work, we investigate a similar set-up, this time considering the holographic duals to a large $N$, strongly coupled CFT on a de Sitter-Schwarzschild background. There are two horizons in the de Sitter-Schwarzschild spacetime, the event horizon and the cosmological horizon, which have different temperatures (except at extremality). The space of such geometries (up to an overall scale) is parameterized by the ratio between the temperatures of the two horizons, or equivalently, by the ratio of the radii of the horizons, $\rho_h = r_h/r_c \in (0,1)$, where $r_h$ and $r_c$ are the proper radii of the event and cosmological horizons, respectively. Throughout this paper, we will refer to the parameter, $\rho_h$, as the \textit{radius ratio}. Note that $\rho_h$ is gauge invariant in this context, due to the background spherical symmetry of the de Sitter-Schwarzschild black hole. This provides a very natural set-up to investigate the phases of the CFT, since rather than having the impose that the CFT is in some thermal state asymptotically, here the geometry naturally imposes that the two horizons radiate with given temperatures.

\begin{figure}[tbh]
    \centering
     \begin{subfigure}{0.7\textwidth}
         \includegraphics[width=\textwidth]{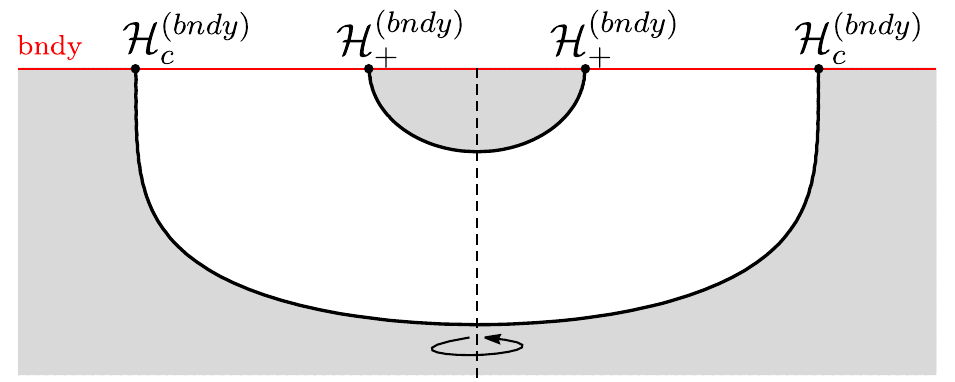}
         \caption{The black tunnel}
         \label{fig:drop_drawing}
     \end{subfigure}
     \par\bigskip
     \centering
     \begin{subfigure}{0.7\textwidth}
         \centering
         \hspace{-30pt}
         \includegraphics[width=\textwidth]{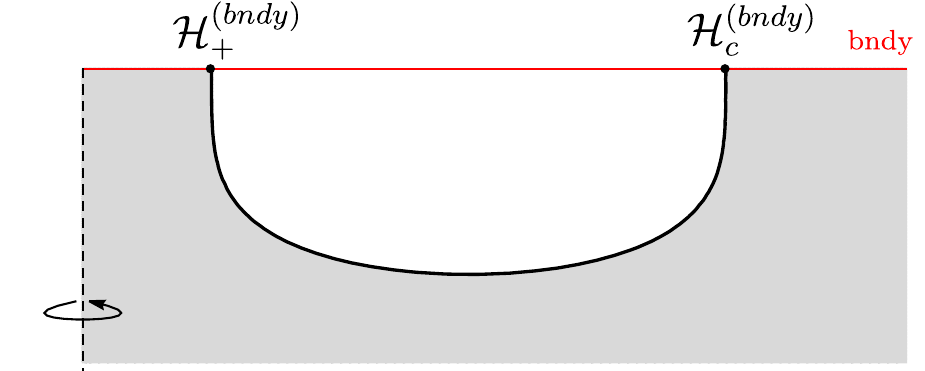}
         \caption{The black hammock}
         \label{fig:ham_drawing}
     \end{subfigure}
     \caption{Some schematic drawings of the black tunnel and hammock at a constant time slice after suppressing two angular directions. In each case the dotted line is the axis of symmetry where the $S^2$ shrinks to zero size, and note that this axis is in slightly different places in the two pictures, so that we've explicitly drawn the $\mathbb{Z}_2$ symmetry for the tunnel, but not the hammock. The hammocks only have one horizon, separating an interior and exterior, which doesn't cross the axis of symmetry, whereas the tunnels have two horizons both of which cross the axis of symmetry. In each diagram, the red line is the conformal boundary, on which there is a de Sitter-Schwarzschild geometry.}\label{fig:drawings}
\end{figure}

Once again, two dual bulk solutions arise, depending on whether the boundary horizons are connected via a bulk horizon or not. The \textit{black tunnel} is a solution in which there are two disconnected bulk horizons, one of which extends from the boundary event horizon and the other from the boundary cosmological horizon. Each of these horizons closes in on itself some way into the bulk. In the other solution, which we call the \textit{black hammock}, the boundary event horizon and boundary cosmological horizons are connected by a single horizon in the bulk. See Figure \ref{fig:drawings} for schematic drawings of these two geometries. Similarly to the black funnels (and other flowing, non-equilibrium steady state solutions \cite{Figueras:2012rb,Emparan:2013fha,Sun:2014xoa,Amado:2015uza,Megias:2015tva,Herzog:2016hob,Megias:2016vae,Sonner:2017jcf,Ecker:2021ukv}), the black hammocks allow classical flow along their horizons, which again is dual to a deconfined phase of the CFT, with $\bigO{N^2}$ Hawking radiation on the field theory side of the duality. There is no classical flow between the two horizons of the black tunnel, so this is dual to a confined phase of the CFT fields with $\bigO{N^0}$ Hawking radiation.

It would be of great interest to investigate which of these solutions dominates for different values of the radius ratio, $\rho_h$, since this would be dual to which phase of the CFT matter is dominant on the field theory side. One would expect a similar phase transition to that of the droplets and funnels, as evidenced by \cite{Santos:2020kmq}. For a number of reasons, which we will discuss, this problem is difficult and may require direct calculation of the stability of the solutions under time-evolution after a small perturbation. This lies outside the scope of this paper, though we hope to return to this question in the near future.

Another point of interest is that the black tunnel and black hammock solutions are closely related to black hole solutions in the Randall-Sundrum II (RSII) model \cite{Randall:1999vf}. Following the methods used in \cite{Figueras:2011gd,Biggs:2021iqw} to find other RSII black hole solutions, one could envisage adapting the black tunnels and hammocks in order to find black hole solutions in the RSII model where one takes a positive effective cosmological constant on the brane. Such solutions would be dual to spherically symmetric four-dimensional black holes with a positive cosmological constant that receive ``quantum corrections'' from a large $N$, strongly coupled CFT.

In this paper, we present the two solutions and the methods used to obtain them. We used different gauge choices in order to solve the Einstein equation numerically for the two different solutions. For the black tunnels, we used \textit{the DeTurck method} \cite{Headrick:2009pv,Wiseman:2011by,Dias:2015nua}. As we'll discuss, this method adds a term to the Einstein equation in order to make the resultant system of PDEs manifestly elliptic, so that they can be solved numerically as a boundary value problem. It turns out in this case that the added term must actually vanish on any solution, hence we still obtain a solution to the Einstein equation.

One could also use the DeTurck method to find the black hammock solutions, and indeed we were able to do so, but we found that we had to use an extremely high number of lattice points and precision in the numerical method in order to extract the quantities of interest from the solutions, making the process extremely computationally expensive. Instead, we found that a different gauge choice, specifically \textit{Bondi-Sachs gauge}, was a lot more effective. Indeed, it seems as though Bondi-Sachs gauge is particularly well-adapted to stationary problems with a null hypersurface in the bulk ``opposite'' the conformal boundary, and so is particularly useful for finding black hammock solutions. The use of the two complementary methods allowed us to find both solutions for a large range of the parameter space $\rho_h \in (0,1)$. To our knowledge this is the first instance where flowing solutions were found by solving a boundary value problem in Bondi-Sachs gauge.

One additional benefit of finding the black hammocks in Bondi-Sachs gauge is that it is a very natural gauge in which to time-evolve the solutions after a slight perturbation. Hence, it provides a way to directly test the stability of the hammocks across the parameter space.

In sections \ref{sec:tunnels} and \ref{sec:hammocks} we present the methods used to find the black tunnels and black hammocks, respectively, and we review how one can extract the holographic stress tensor from the bulk solution in each case. In section \ref{sec:results}, we discuss some of the properties of the solutions. The hammocks have particularly interesting properties due to the fact their horizon is non-Killing. This allows there to be classical flow along the horizon and for the expansion and shear to be non-zero. In section \ref{sec:discussion}, we end with a discussion focusing particularly on the difficulties of deducing which of the black tunnel or black hammock solution, and hence which phase of the dual CFT, dominates for a given value of the radius ratio, $\rho_h$. We argue that in order to obtain the phase diagram of the dual CFT, one would have to investigate directly the stability of the two dual solutions. 

\section{Black tunnels}\label{sec:tunnels}

\subsection{The DeTurck method}
We begin with the method to find the black tunnels. These will be solutions to the Einstein equation in five dimensions with a negative cosmological constant:
\begin{equation}\label{eq:Einstein}
R_{ab}+\frac{4}{L^2}g_{ab}=0\,,
\end{equation}
where the AdS length scale is defined in five dimensions by $L = \sqrt{-6/\Lambda}$, with $\Lambda$ the negative cosmological constant. Due to its diffeomorphism invariance, the Einstein equation does not lead to a well posed problem until a gauge is chosen. For the black tunnels we use \textit{the DeTurck method} \cite{Headrick:2009pv,Wiseman:2011by,Dias:2015nua} to do so. We begin by adding a term on to the Einstein equation to obtain the so-called \textit{Einstein-DeTurck equation} (or harmonic Einstein equation) which is given by
\begin{equation}\label{eq:EDT}
    R_{ab}+\frac{4}{L^2}g_{ab} - \nabla_{(a}\xi_{b)}=0,
\end{equation}
where the DeTurck vector is defined by
\begin{equation}
    \xi^a = g^{cd}\left[\Gamma^a_{cd}(g)-\Gamma^a_{cd}(\Bar{g})\right]
\end{equation}
with $\Gamma^a_{cd}(\mathfrak{g})$ being the Christoffel connection associated to a metric $\mathfrak{g}$, and $\Bar{g}_{ab}$ a reference metric which we can choose freely. The key is that once we take a suitable \textit{Ansatz} for the black tunnels, which is static and spherically symmetric and satisfying certain boundary conditions and, moreover, pick a reference metric that respects these symmetries and boundary conditions, we will find that the Einstein-DeTurck equation yields a system of elliptic PDEs. Such a system can be solved numerically by using pseudo-spectral collocation methods on a Chebyshev-Gauss-Lobatto grid to approximate the PDEs with non-linear algebraic equations, which can then be solved iteratively using Newton's method (see for instance \cite{Dias:2015nua} for a review of these methods in the context of the Einstein equation). We chiefly used a $120\times120$ sized grid in this discretization process.

Of course, solutions to (\ref{eq:EDT}) are only solutions to the Einstein equation if $\xi=0$ on the solution. A solution to the Einstein-DeTurck equation with non-zero $\xi$ is called a \textit{Ricci soliton}. However, it was shown in \cite{Figueras:2011va} that the norm of the DeTurck vector, $\xi^a \xi_a$, obeys a \textit{maximum principle}, which implies that so long as the norm is non-negative across the manifold and vanishing on its boundary, then in fact it must vanish identically across the whole spacetime. If the spacetime is also static, then this disallows the existence of Ricci solitons. We will see that the DeTurck vector satisfies these conditions in the case of the black tunnels, so we can be sure that in our case any solution to the Einstein-DeTurck equation will also be a solution to the Einstein equation.

Moreover, we can keep track of the DeTurck norm of the solutions obtained via this method in order to investigate the accuracy of the numerical methods as one takes more and more points in the discretization. We present such convergence tests for the black tunnels in Appendix \ref{app:tunnels}.

\subsection{Ansatz for the black tunnels}

\begin{figure}[tbh]
    \centering
     \begin{subfigure}{0.472\textwidth}
         \includegraphics[width=\textwidth]{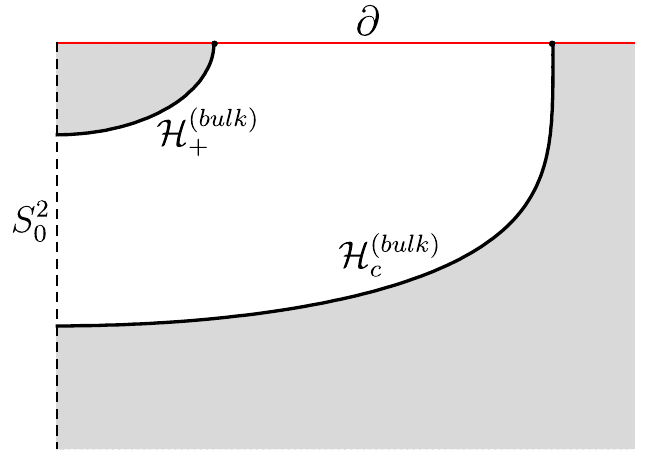}
         \caption{Drawing of the black tunnel}
         \label{subfig:drop_drawing}
     \end{subfigure}
     \centering
     \begin{subfigure}{0.47\textwidth}
         \centering
            \begin{tikzpicture}
            \draw (0,0) -- (0,4) -- (4,4) -- (4,0) -- cycle;
            
            \draw (2,-0.25) node {$S^2_0;\;y=1$};
            \draw (2,4.25) node {$\partial;\;y=0$};
            \draw (-0.7,2.25) node {$\mathcal{H}_+^{(bulk)}$};
            \draw (-0.7,1.75) node {$x=0$};
            \draw (4.7,2.25) node {$\mathcal{H}_c^{(bulk)}$};
            \draw (4.7,1.75) node {$x=1$};
            \end{tikzpicture}

         \caption{Integration domain for the black tunnel}
        
     \end{subfigure}
     \caption{The black tunnel naturally has four boundaries; the conformal boundary ($\partial$), the bulk event horizon ($\mathcal{H}_+^{(bulk)}$), the axis of symmetry where the two-sphere shrinks to zero size ($S^2_0$) and the bulk cosmological horizon ($\mathcal{H}_c^{(bulk)}$). Hence, we automatically have a square integration domain.}\label{fig:drop}
\end{figure}

We assume the solutions will be static and spherically symmetric. On the conformal boundary, which we set at $y=0$, we will enforce the metric to be conformal to that of de Sitter-Schwarzschild. In the bulk there will be two disconnected horizons, one emanating from the event horizon on the boundary and the other from the cosmological horizon. We'll call these the bulk event horizon (at $x=0$) and bulk cosmological horizon (at $x=1$), respectively. Finally, between the two horizons in the bulk, there will be an axis where the $S^2$ shrinks to zero size, given by $y=1$. Hence, the black tunnels naturally live in a rectangular coordinate domain, $\{x,y\} \in (0,1)^2$. In Figure \ref{fig:drop}, we've drawn the tunnel again schematically, along with the integration domain that naturally arises. 

The line element we use to describe such solutions is given by the following \textit{Ansatz}:
\begin{subequations}
\begin{align}\label{eq:drop_Ansatz}
    \diff s^2 = \frac{L^2}{y}\Bigg[&-x^2\left(1-x^2\right)^2 \frac{\left(1-\rho_h\right)^2}{\rho_h^2}G(x)q_1(x,y)\diff T^2 + \frac{16}{G(x)}q_2(x,y)\diff x^2 \nonumber \\
    &\; +(1-y)q_3(x,y)\diff \Omega^2_{(2)} + \frac{q_4(x,y)\left(\diff y+(1-y)q_5(x,y)\diff x\right)^2}{4(1-y)y}\Bigg],
\end{align}
where
\begin{equation}\label{eq:G_defn}
    G(x)=\frac{(2-x^2)\left(1+2\rho_h-x^2(2-x^2)(1-\rho_h^2)\right)}{1+\rho_h+\rho_h^2}\,,
\end{equation}
\end{subequations}
and $\diff \Omega^2_{(2)}$ is the metric on a round, unit radius two-sphere. The parameter, $\rho_h$, is the ratio of the radii of the boundary event and boundary cosmological horizon in the de Sitter-Schwarzschild geometry which we will enforce is conformal to the induced metric on the conformal boundary. Hence $\rho_h \to 1$ is the extremal limit at which the event and cosmological horizons are coincident, whilst $\rho_h \to 0$ is the limit where the horizons become infinitely far apart. We found black tunnel solutions across the whole of the parameter space $\rho_h \in (0,1)$.

The reference metric we used to define the DeTurck vector is given by the \textit{Ansatz} above with $q_1=q_2=q_3=q_4=1$ and $q_5=0$.

\subsection{Boundary conditions}
The Einstein-DeTurck equation evaluated for the \textit{Ansatz} given by (\ref{eq:drop_Ansatz}) yields a system of five second-order PDEs on the integration domain $\{x,y\} \in (0,1)^2$. In order to solve such a system, we need to set five boundary conditions on each of the sides of the square. 
\subsubsection{The conformal boundary $y=0$}
Here, we enforce Dirichlet boundary conditions in order to set the induced metric on the conformal boundary to be conformal to the de Sitter-Schwarzschild metric in four dimensions, which is given by
\begin{equation}\label{eq:dS-Schw_r}
    \diff s^2_{dS-S} = -f(r)\diff t^2+\frac{\diff r^2}{f(r)} + r^2\diff \Omega^2_{(2)},
\end{equation}
where
\begin{align}
    f(r) &= 1 - \frac{2M}{r}-\frac{\Lambda_4 r^2}{3} \nonumber \\
    &= -\frac{(r-r_c)(r-r_h)(r+r_c+r_h)}{r\left(r_c^2+r_c r_h + r_h^2\right)},
\end{align}
with $\Lambda_4$ the positive cosmological constant of the four-dimensional geometry, $M$ the mass of the black hole and $r_h$ and $r_c$ the radii of the event and cosmological horizons, respectively. Taking the transformations
 \begin{equation}
     r = \frac{r_h}{1-\xi^2(2-\xi^2)(1-\rho_h)},\quad \text{with}\quad \rho_h = \frac{r_h}{r_c},
 \end{equation}
 the metric becomes
 \begin{equation}\label{eq:dS-Schw_xi}
     \diff s^2_{dS-S} = \frac{r_h^2}{\left(1-\xi^2(2-\xi^2)(1-\rho_h)\right)^2}\left[-\xi^2(1-\xi^2)^2 \frac{\left(1-\rho_h\right)^2}{\rho_h^2 r_c^2}G(\xi)\diff t^2 + \frac{16 \diff \xi^2}{G(\xi)}+\diff\Omega^2_{(2)}\right],
 \end{equation}
 where $G$ is defined by (\ref{eq:G_defn}). In these coordinates, the event horizon lies at $\xi=0$ and the cosmological horizon is situated at $\xi=1$. With respect to the usual $t$-coordinate, the temperature of the event and cosmological horizons are, respectively, given by
 \begin{equation}\label{eq:dS_temp}
     T_H = \frac{(1-\rho_h)(1 + 2\rho_h)}{4\pi r_h \left(1+\rho_h+\rho_h^2\right)}, \quad 
     T_c =  \frac{(1-\rho_h)(2 + \rho_h)}{4\pi r_c \left(1+\rho_h+\rho_h^2\right)}.
 \end{equation}
 Now we are ready to define our boundary conditions at $y=0$, the conformal boundary. Here, we set $q_1=q_2=q_3=q_4=1$, and $q_5=0$. With such a choice, and taking
 \begin{equation}
     y = \frac{\left(1-\xi^2(2-\xi^2)(1-\rho_h)\right)^2}{r_h^2}z^2, \quad x = \xi, \quad T = \frac{t}{r_c},
 \end{equation}
 one finds that at leading order in $z$, near the conformal boundary
 \begin{equation}
     \diff s^2 \Big{|}_{z=0} = \frac{L^2}{z^2}\left(\diff z^2 + \diff s^2_{dS-S}\right),
 \end{equation}
 with the de Sitter-Schwarzschild metric being given by (\ref{eq:dS-Schw_xi}). Hence, this boundary condition enforces that the black tunnel is an AlAdS spacetime with the metric on the conformal boundary being conformal to de Sitter-Schwarzschild.
 
\subsubsection{The fictitious boundaries}
The remaining three boundaries of the integration domain are fictituous boundaries where derived boundary conditions are obtained \cite{Headrick:2009pv,Wiseman:2011by,Dias:2015nua}. At $x=0$ and $x=1$, respectively, we have the bulk event and cosmological horizons and at $y=1$ we have the axis between the two bulk horizons where the two-sphere collapses to zero size.

At each of these boundaries we require that the metric is regular. This can be imposed with Neumann boundary conditions. In particular, at each fictitious boundary we set the normal derivative of each function to the boundary to zero. As an example, at $x=0$, we set
\begin{equation}
    \partial_x q_i(0,y)=0, \quad \text{for} \; i=1,\hdots, 5.
\end{equation}

\subsection{Extracting the holographic stress tensor}\label{sec:drop_stress}
Now let us briefly discuss how we can extract the holographic stress tensor once we have obtained the solution numerically. Firstly, one can solve the equations of motion defined by $(\ref{eq:EDT})$ order by order in $y$ off the boundary $y=0$. This fixes that
 \begin{align}\label{eq:drop_expansion}
     q_1(x,y)&=1+\alpha_1(x)y+\beta_1(x)y^2 + \hat{\gamma}_1(x) y^{1+\sqrt{3}} + \ldots \nonumber\\
     q_2(x,y)&=1+\alpha_1(x)y+\beta_2(x)y^2 + \hat{\gamma}_1(x) y^{1+\sqrt{3}} +\ldots\nonumber\\
     q_3(x,y)&=1-\inv{2}\alpha_1(x)y+\beta_3(x)y^2 + \hat{\gamma}_1(x) y^{1+\sqrt{3}} + \ldots\nonumber\\
     q_4(x,y)&=1+\beta_4(x)y^2 + \hat{\gamma}_4(x) y^{1+\sqrt{3}} +  \ldots\nonumber\\
     q_5(x,y)&=\beta_5(x)y^2+\gamma_5(x)y^3 + \Tilde{\gamma}_5(x)y^3 \log{y} + \ldots
 \end{align}
 Some of these functions are fixed by the equations of motion as
\begin{subequations}
 \begin{equation}
     \alpha_1(x)=-\frac{1+\rho _h}{1+\rho _h+\rho _h^2} g(x)
     \end{equation}
     \begin{equation}
    \beta_3(x)+\frac{\beta _1(x)}{2}+\frac{\beta _2(x)}{2}=\frac{5 (1+\rho _h)^2 g(x)^2 - 2 (1+\rho _h) (1+\rho _h+\rho _h^2) g(x)}{8\left(1+\rho _h+\rho _h^2\right)^2}
    \end{equation}
    \begin{equation}
    \beta_4(x)=\frac{2 (1+\rho _h+\rho _h^2) (1+\rho _h) g(x)-(1+\rho_h)^2 g(x)^2}{4\left(1+\rho _h+\rho _h^2\right)^2}
    \end{equation}
    \begin{equation}
     \beta_5(x)=-\frac{(1+\rho _h)}{2 (1+\rho _h+\rho _h^2)} g^\prime(x)
     \end{equation}
     \begin{equation}
     \Tilde{\gamma}_5(x)=\frac{3 (1+\rho _h)^2}{8 (1+\rho _h+\rho _h^2)^2}g(x) g^\prime(x)\,,
 \end{equation}
 where
\begin{equation}
g(x)=1-x^2(2-x^2)(1-\rho_h)\,.
\end{equation}
 \end{subequations}
The only functions not fixed by a local analysis of the Einstein-DeTurck equation off the conformal boundary are $\{\beta_1, \beta_2, \gamma_5, \hat{\gamma}_1, \hat{\gamma}_4\}$. In order to find these functions, we need to solve the equations in the full spacetime, after having imposed regularity deep in the bulk. It turns out only the $\beta_i$ functions are needed to calculate the stress tensor. Note that, once we have numerical approximations of the full functions $q_i$, we can easily evaluate an estimate for $\beta_i$ from the second derivative of $q_i$ with respect to $y$ at $y=0$.
 
Armed with this expansion near $y=0$, we can go to Fefferman-Graham gauge \cite{Fefferman:1985} near the conformal boundary and fix the conformal frame. That is, we seek a coordinate transformation such that near the conformal boundary the metric takes the form
\begin{equation}\label{eq:FG}
    \diff s^2=\frac{L^2}{z^2}\left[\diff z^2+ \left(\Bar{g}_{\mu\nu}+A_{\mu\nu} z^2+ B_{\mu\nu}z^4+ C_{\mu\nu}z^4\log{z}\right)\diff x^\mu \diff x^\nu + \bigO{z^5}\right],
\end{equation}
where we pick the conformal frame so that $\Bar{g}_{\mu\nu}$ is the de Sitter-Schwarzschild metric given by (\ref{eq:dS-Schw_xi}). This can be achieved with a transformation 
\begin{align}
    x &= \xi+\sum_{j=1}^6 \delta_j(\xi) z^j \nonumber\\
    y &= \sum_{j=2}^6 \epsilon_j(\xi) z^j.
\end{align}
The explicit expressions of $\delta_j(\xi)$ and $\epsilon_j(\xi)$ can be determined by substituting the above transformation into the metric given by the \textit{Ansatz} after expanding each of the functions off the boundary with (\ref{eq:drop_expansion}). One can then work order by order in $z$ to match the resultant metric with (\ref{eq:FG}). Such a procedure fixes $A_{\mu\nu},\,B_{\mu\nu}$ and $C_{\mu\nu}$ uniquely. In our case we find $C_{\mu\nu}=0$, in accordance with the fact that our boundary metric is Einstein and thus the field theory in this case should be anomaly free \cite{deHaro:2000vlm}.

With the metric in Fefferman-Graham coordinates, one can readily read off the holographic stress tensor \cite{deHaro:2000vlm}, by calculating
\begin{align}\label{eq:drop_stress}
    \langle T_{\mu\nu}\rangle = \frac{L^3}{4\pi G_5}\left[B_{\mu\nu} + \inv{8}\left(A_{\rho\sigma}A^{\rho\sigma} - (\Tr A)^2 \right)\Bar{g}_{\mu\nu} - \inv{2}A_{\mu\rho}A^{\rho}{}_{\nu} + \inv{4}A_{\mu\nu}\Tr A \right],
\end{align}
where indices are risen and lowered with respect to the boundary metric $\Bar{g}_{\mu\nu}$, and hence the trace is given by $\Tr A = A^{\mu \nu}\Bar{g}_{\mu\nu}$. Finally, we can make the identification from the standard AdS/CFT dictionary that
\begin{equation}
    G_5 = \frac{\pi}{2}\frac{L^3}{N^2}.
\end{equation}
One can check that the stress tensor is conserved and has a fixed trace:
\begin{equation}\label{eq:stress_trace}
    \nabla_\mu T^{\mu\nu} = 0,\quad T^\mu{}_\mu = -\frac{3L^3}{16 \pi G_5 r_c^4\left(1+\rho_h+\rho_h^2\right)^2} = -\frac{3 L^3}{16 \pi G_5 \ell_4^4},
\end{equation}
where $\ell_4 = \sqrt{3/\Lambda_4}$ is the de Sitter length scale of the four-dimensional boundary geometry.

After the dust settles, we find
\begin{subequations}
\begin{multline}
\langle T^{t}_{\phantom{t}t}\rangle=\frac{N^2 g(\xi )^4}{2 \pi ^2 r_c^4 \rho _h^4} \Bigg\{\beta _1(x)
-\frac{1}{16 (1+\rho _h+\rho _h^2)^2}\Bigg[\frac{3 \rho _h^4}{g(\xi )^4}+\frac{4 \rho _h^2 (1+\rho _h)}{g(\xi )}
\\
-12 (1+\rho_h)(1+\rho_h+\rho_h^2) g(\xi )+12 (1+\rho
   _h)^2 g(\xi )^2\Bigg]\Bigg\}\,,
\end{multline}
\begin{multline}
\langle T^{\xi}_{\phantom{\xi}\xi}\rangle=\frac{N^2 g(\xi )^4}{2 \pi ^2 r_c^4 \rho _h^4} \Bigg\{\beta _2(x)
-\frac{1}{16 (1+\rho _h+\rho _h^2)^2}\Bigg[\frac{3 \rho _h^4}{g(\xi )^4}
\\
-8 (1+\rho_h)(1+\rho_h+\rho_h^2) g(\xi )+8 (1+\rho
   _h)^2 g(\xi )^2\Bigg]\Bigg\}\,,
\end{multline}
\begin{multline}
\langle T^{\Omega_i}_{\phantom{\Omega_i}\Omega_i}\rangle=\frac{N^2 g(\xi )^4}{2 \pi ^2 r_c^4 \rho _h^4} \Bigg\{\beta _3(x)
-\frac{1}{16 (1+\rho _h+\rho _h^2)^2}\Bigg[\frac{3 \rho _h^4}{g(\xi )^4}-\frac{2 \rho _h^2 (1+\rho _h)}{g(\xi )}
\\
+6 (1+\rho_h)(1+\rho_h+\rho_h^2) g(\xi )\Bigg]\Bigg\}\,,
\end{multline}
\end{subequations}
where $\Omega_i$ stands for any of the angles on the round $S^2$.

\section{Black hammocks}\label{sec:hammocks}
In order to obtain the black hammocks we use a different method to the DeTurck trick used to find the black tunnels. Instead, we solve the Einstein equation in Bondi-Sachs gauge. We found that this choice of gauge allowed far easier computation of the black hammock solutions and their properties than the DeTurck gauge, and this opens up the possibility that it may be an excellent choice in order to solve similar problems.

We will first consider the integration domain of the black hammocks, and review a standard trick used in order to attain a square domain \cite{Santos:2012he}, which is easier to carry out the numerics on. We'll then review the Bondi-Sachs gauge in general and the scheme used to solve Einstein's equations, before returning to the specific problem regarding the hammocks by presenting the \textit{Ansatz} for them in Bondi-Sachs gauge and explaining the boundary conditions and how some properties of the solutions can be extracted. 
\subsection{The integration domain for the black hammocks}

\begin{figure}[tb]
    \centering
    \hspace{-50pt}
     \begin{subfigure}{0.6\textwidth}
        \hspace{-20pt}
         \includegraphics[width=\textwidth]{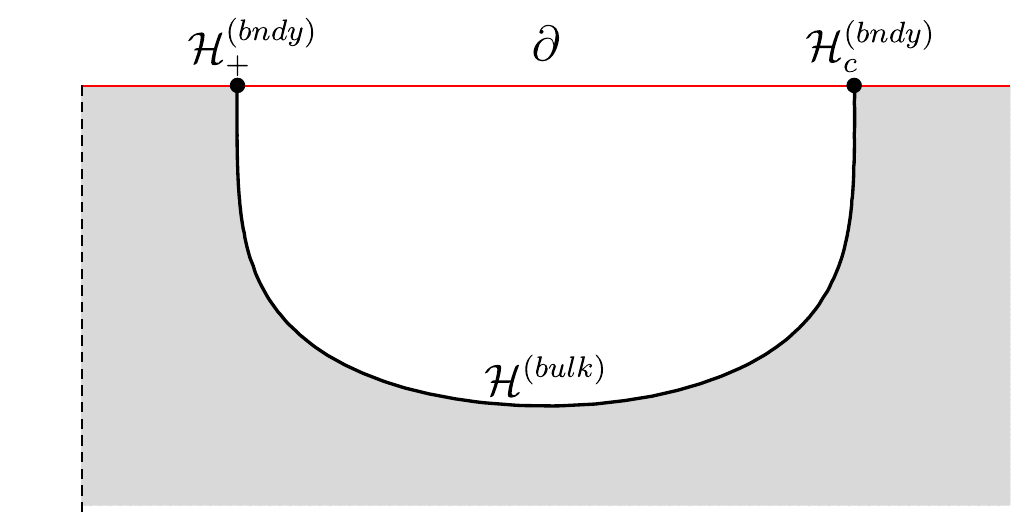}
         \caption{Drawing of the black hammock}
         \label{fig:ham_drawing_2}
     \end{subfigure}
     \centering
     \begin{subfigure}{0.3\textwidth}
            
            \begin{tikzpicture}
            \draw (0,0) -- (0,4) -- (4,4) -- (4,0) -- cycle;
            
            \draw (2,-0.25) node {$\mathcal{H}^{(bulk)};\;y=1$};
            \draw (2,4.25) node {$\partial;\;y=0$};
            \draw (-0.4,2.25) node {$\mathbb{H}_+$};
            \draw (-0.6,1.75) node {$x=0$};
            \draw (4.4,2.25) node {$\mathbb{H}_c$};
            \draw (4.6,1.75) node {$x=1$};
            \end{tikzpicture}
         \caption{Integration domain for the black hammock}\label{fig:ham_domain}
     \end{subfigure}
     \caption{Here we sketch the black hammock in \textbf{(a)}. It initially appears to have only two boundaries: the conformal boundary and the bulk horizon. However, the bulk horizon must approach hyperbolic black holes as it approaches the conformal boundary, where it meets the boundary event horizon and the boundary cosmological horizon. This allows us one to add two extra boundaries to the integration domain at $x=0$ and $x=1$, where this limiting behaviour will be imposed. Hence, we obtain a rectangular integration domain as shown in \textbf{(b)}.}.\label{fig:ham}
\end{figure}

At first sight, it seems as though the black hammocks will only have two boundaries: the conformal boundary and the horizon of the bulk black hole, which now hangs down in the bulk between the positions of the boundary event horizon and the boundary cosmological horizon, motivating the name ``hammock''. We've schematically drawn the shape of this geometry in Figure \ref{fig:ham_drawing_2}.

Note that these black hammocks share many similarities with the black funnel solutions of \cite{Santos:2012he,Fischetti:2016oyo,Santos:2020kmq}, which were gravitational duals to a CFT living on a flat Schwarzschild background. The key difference is the horizons of such black funnel solutions approach a planar black hole in the bulk far from the boundary event horizon, due to the fact the flat Schwarzschild boundary metric possesses only one horizon. On the other hand, the black hammocks will have no such asymptotic region; the bulk horizon will return to the conformal boundary to intersect the boundary cosmological horizon.

Attempting naively to carry out numerics on a domain with only two boundaries would be very difficult, however, fortunately we can use a trick to ``blow up'' the points at which the horizon meets the boundary into lines, as was done in \cite{Santos:2012he, Fischetti:2012vt,Fischetti:2016oyo,Santos:2020kmq}, due to the fact that the bulk horizon must approach a hyperbolic black hole as it approaches the boundary on either side. The reason behind this is that, at leading order, the geometry near the conformal boundary must be hyperbolic, and hyperbolic black holes are the only family of static geometries which are manifestly hyperbolic for each time slice as one approaches the boundary. This family runs over one parameter: the temperature of the hyperbolic black hole. It is likely that one could find \textit{detuned} black hammock solutions, that is, solutions where the temperature of the hyperbolic black hole is different to the temperature of boundary horizon that the bulk horizon is approaching. Such detuned solutions were found in the context of field theories living on an asymptotically flat Schwarzschild black hole \cite{Fischetti:2016oyo}. In this work however, we choose not to study the solutions in which the temperatures of the hyperbolic black holes match the temperatures of the horizons of the boundary de Sitter-Schwarzschild geometry.

The metric of such a hyperbolic black hole is given by 
\begin{equation}\label{eq:hyp}
    \diff s^2_{\mathbb{H}}=\frac{L^2}{z^2}\left[-(1-z^2)\diff \hat{t} +\frac{\diff z^2}{1-z^2} +\diff \eta^2 + \sinh^2\eta \diff \Omega^2_{(2)}\right],
\end{equation}
which has temperature $(4\pi)^{-1}$ with respect to the time coordinate $\hat{t}$.

By blowing up the point where the bulk horizon meets the conformal boundary into lines, we obtain a rectangular integration domain, as shown in Figure \ref{fig:ham_domain}. We'll pick coordinates such that the boundaries  are the conformal boundary at $y=0$, the two hyperbolic black hole limits at $x=0$ and $x=1$ and finally the bulk horizon at $y=1$.

\subsection{Bondi-Sachs gauge}\label{sec:BS}
We wish to specify an \textit{Ansatz} that satisfies the desired symmetries and boundary conditions of the black hammock and then solve the Einstein equation to find the metric. However, as discussed in section \ref{sec:tunnels}, the diffeomorphism invariance of the Einstein equation causes them to lead to a set of mixed hyperbolic-elliptic PDEs, which are difficult to solve numerically. Once again, this can be alleviated by picking a gauge. Rather than using the DeTurck trick, as we did for the black tunnels, we found that one particularly useful gauge was Bondi-Sachs gauge in AdS. This gauge was discussed in references \cite{Balasubramanian:2013yqa,Poole:2018koa, Compere:2019bua} in four dimensions, and here we extend the results to the very similar case of five dimensions with spherical symmetry. This gauge seems to be particularly well suited for the problem at hand, where there is a null hypersurface (in this case the bulk horizon) opposite the conformal boundary in the integration domain. We were also able to find the black hammocks in the DeTurck gauge, however we found that the numerics were far quicker and more accurate working in Bondi-Sachs gauge. Let us discuss this gauge in general here and then in the next subsection we will focus on how it can be used for the black hammocks specifically.

For a general AlAdS metric in five dimensions, with conformal boundary at $y=0$, the metric can be locally written in Bondi-Sachs gauge as
\begin{equation}
    \diff s^2 = \frac{L^2}{y^2}\Big[e^{2\beta}\left(-V \diff v^2 - 2\diff v \diff y\right) + e^{2\chi}h_{IJ}\left(\diff x^I - U^I \diff v\right)\left(\diff x^J - U^J \diff v\right) \Big],
\end{equation}
where $I,J=1,2,3$ for a five-dimensional spacetime, and we've taken the determinant of $h_{IJ}$ into the function $\chi$, so that $\det{h}=1$. We are interested in a spherically symmetric solution, in which case the metric can be written as
\begin{equation}\label{eq:BS_gauge}
    \diff s^2 = \frac{L^2}{y^2}\left[e^{2\beta}\left(-V \diff v^2 - 2\diff v \diff y\right) + e^{2\chi}\left(\inv{A^2}\left(\diff x - U^x \diff v\right)^2+ A\diff \Omega^2_{(2)}\right)\right],
\end{equation}
with five functions, $V,\,U^x,\,\beta,\,A,\,\chi$, which can depend upon $\{v,x,y\}$. For a stationary solution, these functions are independent of $v$. Note that by redefining the radial coordinate, $y$, we also have the freedom to fix one of $\beta$ or $\chi$ (or a combination thereof).

In the next subsection we will present the \textit{Ansatz} for the black hammocks in Bondi-Sachs gauge. We will then solve the Einstein equation with negative cosmological constant:
\begin{equation}\label{eq:ham_Einstein}
    E_{ab}:= R_{ab}+\frac{4}{L^2}g_{ab}=0.
\end{equation} 
However, the Einstein equation has more non-trivial components than the number of free functions, $\{V,\,U^x,\,\beta,\,A,\,\chi\}$, but it so happens that not all of these components are independent. Following \cite{Balasubramanian:2013yqa,Poole:2018koa, Compere:2019bua}, it turns out that one need only solve the four independent equations of motion coming from $E_{ij} = 0$, with $i,j \neq v$ in the bulk - we call these \textit{the bulk equations}. This leaves three other non-trivial equations, $E_{va} = 0$, which must also be satisfied in order to have a full solution to the Einstein equation. However, it can be shown using the contracted Bianchi identity that if these additional equations are satisfied at some constant $y$ slice, then actually they are necessarily satisfied throughout the whole of the bulk, so long as the bulk equations are also satisfied there.
The contracted Bianchi identity is given by
\begin{equation}
    \nabla_a R^a{}_b = -\inv{2}\nabla_b R.
\end{equation}
We'll work in the $\{v,x,y,\theta,\phi\}$ coordinates and here Greek indices will run over all components, $i,\,j$ indices run over $\{x,y,\theta,\phi\}$, and $A,\,B$ indices run over $\{x,\theta,\phi\}$. In a coordinate basis, the Bianchi identity can be expanded to the equality
\begin{equation}\label{eq:Bianchi}
    0 = g^{\mu\rho}\left(\partial_\mu R_{\nu\rho}-\inv{2}\partial_\nu R_{\mu\rho}-\Gamma^\sigma_{\mu\rho}R_{\nu\rho}\right).
\end{equation}
We'll assume that the bulk equations hold, \textit{i.e.} we have
\begin{equation}
    R_{yy}=0,\quad R_{yx}=0,\quad R_{xx} =- \frac{4}{L^2}g_{xx},\quad R_{\phi\phi} = -\frac{4}{L^2}g_{\phi\phi},
\end{equation}
where we've used $\theta$ and $\phi$ as the usual coordinates on the two-sphere. Let us first consider the $y$-component of the contracted Bianchi identity, given in (\ref{eq:Bianchi}). Noting that the inverse metric has $g^{v\mu} = 0$ for all $\mu \neq y$, and after applying the bulk equations, one finds that the $y$-component of the contracted Bianchi identity simplifies to
\begin{align}\label{eq:y_Bianchi}
    0 &= \frac{2}{L^2}g^{AB}\partial_y g_{AB} - g^{\nu\rho}\Gamma^v_{\nu\rho}R_{yv}.
\end{align}
One can directly calculate the inverse metric and the Christoffel symbols, and find that (\ref{eq:y_Bianchi}) is solved by
\begin{equation}\label{eq:Eyv}
    R_{yv} = -\frac{4}{L^2}g_{yv}.
\end{equation}
Thus the bulk equations automatically solve $E_{yv}=0$ algebraically. 

It still remains to satisfy the $(v,v)$ and the $(v,x)$ components of the Einstein equation. For the latter, consider the $x$-component of the contracted Bianchi identity. We have
\begin{equation}
    0 = g^{\mu\rho}\left(\partial_\mu R_{x\rho}-\inv{2}\partial_x R_{\mu\rho}-\Gamma^\sigma_{\mu\rho}R_{x\rho}\right).
\end{equation}
Expanding all of these terms, applying the bulk equations, as well as (\ref{eq:Eyv}) yields
\begin{align}
    g^{vy}\partial_y R_{vx} - g^{\mu\rho}\Gamma^v_{\mu\rho}R_{vx} = \frac{4}{L^2}\Bigg[-g^{vy}\partial_x &g_{vy} + g^{xy}\partial_y g^{vy} + g^{xy}\partial_y g_{xx} + \inv{2}g^{xx}\partial_x g_{xx} \nonumber \\
    &-\inv{2}g^{\theta\theta}\partial_x g_{\theta\theta}-\inv{2}g^{\phi\phi}\partial_x g_{\phi\phi} - g^{\mu\rho}\Gamma^x_{\mu\rho}g_{xx}\Bigg].
\end{align}
Now by direct calculation, one can check that $R_{vx} = -(4/L^2)g_{vx}$, or equivalently $E_{vx}=0$, solves the above equation. Since the equation is a first-order PDE, possessing only a $y$-derivative, this means that so long as we set $E_{vx}=0$ on some constant $y$ slice, then the contracted Bianchi identity ensures that the unique solution must satisfy $E_{vx}=0$ throughout the whole bulk, via Picard's theorem.

We can deal with the $(v,v)$ component of the Einstein equation in a similar fashion by considering the $v$-component of the contracted Bianchi identity:
\begin{equation}
    0 = g^{\mu\rho}\left(\partial_\mu R_{v\rho}-\inv{2}\partial_v R_{\mu\rho}-\Gamma^\sigma_{\mu\rho}R_{v\rho}\right).
\end{equation}
Once again, we expand each term, apply the bulk equations, (\ref{eq:Eyv}), and now additionally $E_{vx}=0$. This gives
\begin{align}
    g^{vy}\partial_y R_{vv} - g^{\mu\rho}\Gamma^v_{\mu\rho}R_{vv} = \frac{4}{L^2}\Bigg[&g^{yy}\partial_y g_{vy} + g^{yx}\left(\partial_y g_{vx} +\partial_x g_{vy}\right) + g^{xx}\left(\partial_x g_{vx} - \inv{2}\partial_v g_{xx}\right) \nonumber \\
    &-\inv{2}g^{\theta\theta}\partial_vg_{\theta\theta}-\inv{2}g^{\phi\phi}\partial_v g_{\phi\phi} - g^{\mu\rho}\left(\Gamma^y_{\mu\rho}g_{vy}+\Gamma^x_{\mu\rho}g_{vx}\right)\Bigg].
\end{align}
Again by direct calculation, one can verify that $R_{vv} = -(4/L^2)g_{vv}$ solves this equation, hence by a similar argument to the $(v,x)$ component, so long as we enforce $E_{vv} = 0$ on a constant $y$ slice, it will be satisfied throughout the whole bulk.

To summarise, when working with a spherically symmetric metric in Bondi-Sachs gauge, given by (\ref{eq:BS_gauge}), one need not solve each and every component of the Einstein equation across the whole space in order to get a solution. Instead, one can solve the four bulk equations coming from $E_{ij} = 0$ for $i,j \neq v$, and additionally set $E_{va} = 0$ at some constant $y$ hypersurface, say as boundary conditions, and then the contracted Bianchi identity enforces that actually $E_{va} = 0$ throughout the whole spacetime. In this way, one can attain an Einstein metric. This is the approach we take to find the metric of the black hammocks.

One way to monitor the convergence of our numerical method is to monitor $E_{va}$ throughout spacetime and check that it remains zero away from where we imposed $E_{va} = 0$ as a boundary condition. In this way, the method we outline here is not very different from the DeTurck method, where we also have to check that $\xi_a$ vanishes in the continuum limit.
\subsection{Ansatz for the black hammocks}

For the black hammocks, we take the \textit{Ansatz} to be
\begin{align}\label{eq:ham_Ansatz}
    \diff s^2 = \frac{L^2}{y^2}\Bigg[&\left(-(1-y^2)\frac{(1-\rho_h)^2}{\rho_h^2}H(x)^2 p_1(x,y) \diff v^2 - 2\frac{(1-\rho_h)}{\rho_h}H(x)\diff v \diff y\right)p_3(x,y)^2 \nonumber \\
    & \inv{4x(1-x)\left(1+y^4p_5(x)\right)^2}\left(\frac{\left(\diff x - x(1-x)p_2(x,y)\diff v\right)^2}{(1-x)x\, p_4(x,y)^2}+H(x)p_4(x,y)\diff \Omega^2_{(2)}\right) \Bigg],
\end{align}
where
\begin{equation}\label{eq:Heqn}
    H(x) = \frac{1+2\rho_h-x(1-\rho_h^2)} {1+\rho_h+\rho_h^2}.
\end{equation}
After a few redefinitions, one can see that indeed this metric is in Bondi-Sachs gauge (\ref{eq:BS_gauge}), where we've used the freedom of redefining the radial coordinate to fix the radial dependence of the $\chi$ function. Any metric defined by this \textit{Ansatz} clearly contains a null hypersurface at $y=1$, is spherically symmetric and possesses another Killing vector, $\partial / \partial v$, which we'll see is timelike outside of an ergoregion that is located near the $y=1$ null hypersurface.

The bulk equations, $E_{ij}=0$, provide four PDEs for the functions $p_k(x,y)$ for $k=1,\hdots,4$ and $p_5(x)$. It's worth noting that $p_5$ is independent of $y$, and moreover, the $E_{xy} = 0$ equation evaluated at $y=1$ actually provides a simple algebraic condition for $p_5(x)$ in terms of the four other functions and their derivatives at $y=1$. However, we found that the convergence properties of the numerical method was faster if we promoted $p_5$ to a function of both $x$ and $y$ and imposed as a fifth bulk equation that $\partial_y p_5(x,y) = 0$.

The highest $y$-derivative of the functions arising in the bulk equations are
\begin{equation}
\{\partial_y p_1, \partial^2_y p_2,\partial_y p_3, \partial^2_y p_4\}\,,
\end{equation}
that is, two second order derivatives and two first order derivatives with respect to $y$. This suggests that when we expand the functions about the conformal boundary, we should find there are three free functions in order for us to have a well-defined PDE problem. Moreover, we should find we need to set three boundary conditions deep in the bulk at $y=1$.

We will enforce that $E_{va} = 0$ at the $y=1$ hypersurface. As discussed above, the contracted Bianchi identity implies that, so long as the bulk equations are satisfied, such boundary conditions actually enforce that $E_{va} = 0$ throughout the whole bulk, yielding a full solution to the Einstein equation. Indeed, once a solution is found we can check explicitly that the $E_{va}$ components vanish (within our numerical error), providing a test of the convergence properties of the numerical method to find the black hammocks, which we present in Appendix \ref{app:hammocks}.

In order to numerically solve the bulk equations, we must firstly set boundary conditions, which we discuss in the next subsection, and then, once again, we use collocation methods and Newton's method to numerically solve the system of PDEs obtained. For the black hammocks, we used a grid of size $120\times120$ when discretizing.

It's worth noting that we have no proof that the system of PDEs that result from the Einstein equation applied to our \textit{Ansatz} in Bondi-Sachs gauge is elliptic, though the speed and accuracy of the numerical method would seem to suggest ellipticity. It would be very interesting to explore further whether this gauge naturally gives elliptic PDEs for such stationary problems in general relativity and we will leave this discussion to future work.
\subsection{Boundary conditions}

\subsubsection{The conformal boundary $y=0$}
Just as with the black tunnels, at the conformal boundary we want to set Dirichlet boundary conditions enforcing that the metric on the conformal boundary is conformal to the de Sitter-Schwarzschild metric, given by (\ref{eq:dS-Schw_r}). This time, consider a coordinate transformation given by 
\begin{equation}
    \quad r = \frac{r_h}{1-\left(1-\rho_h\right)w}, \quad \text{with} \quad \rho_h=\frac{r_h}{r_c},
\end{equation}
so that the event horizon is the $w=0$ surface and the cosmological is the $w=1$ surface, and the metric is given by
 \begin{equation}\label{eq:dS-Schw_w}
     \diff s^2_{dS-S} = \frac{r_h^2}{\left(1-w(1-\rho_h)\right)^2}\left[-w(1-w) \frac{\left(1-\rho_h\right)^2}{\rho_h^2 r_c^2}H(w)\diff t^2 + \frac{\diff w^2}{H(w)w(1-w)}+\diff\Omega^2_{(2)}\right],
 \end{equation}
 where $H$ is defined by (\ref{eq:Heqn}). We set at $y=0$ the boundary conditions that $p_1=p_3=p_4=1$ and $p_2=0$, and then take the transformations
 \begin{equation}\label{eq:ham_yM_trans}
     v = \frac{t}{2 r_c} - \frac{\rho_h}{H(w)(1-\rho_h)}y, \quad y = z \frac{(1-w(1-\rho_h))\sqrt{H(w)}}{2 r_h \sqrt{w(1-w)}}, \quad x=w,
 \end{equation}
 then at leading order in $z$, the metric near the $z=0$ boundary is
 \begin{equation}
     \diff s^2 = \frac{L^2}{z^2}\left(\diff z^2 +\diff s^2_{dS-S}\right),
 \end{equation}
 with the de Sitter-Schwarzschild metric in the coordinates of (\ref{eq:dS-Schw_w}). Therefore, the black hammocks will also be an AlAdS spacetime with de Sitter-Schwarzschild on the boundary.
 
\subsubsection{The hyperbolic black hole $x=0$}
As discussed above, we use the fact that the bulk horizon must approach the geometry of the horizon of a hyperbolic black hole as it approaches the conformal boundary to add another boundary to the integration domain. In order to enforce that the metric takes the form (\ref{eq:hyp}) at this boundary, $x=0$, we take the Dirichlet boundary conditions: $p_1=p_3=p_4=1$ and $p_2=0$. Moreover, we also set $p_5=0$ at $x=0$, though it can be easily shown that this follows from the equations of motion. Now, if we take new coordinates
\begin{equation}
    \diff v = \inv{r_c}\diff \hat{t} - \frac{\rho_h}{H(0)(1-\rho_h)(1-y^2)}\diff{y}, \quad x = H(0)\xi,
\end{equation}
then at leading order in $\xi$ for each component, the metric becomes
\begin{equation}
    \diff s^2\Big|_{\xi=0} = \frac{L^2}{y^2}\left[-\frac{(1-y^2)(1-\rho_h)^2 H(0)^2}{\rho_h^2 r_c^2}\diff \hat{t}^2 + \frac{\diff y^2}{1-y^2}+ \inv{4\xi^2}\diff \xi^2+\inv{4\xi}\diff\Omega^2_{(2)} \right].
\end{equation}
After taking 
\begin{equation}
    \xi = e^{-2\eta}, \quad y = z,
\end{equation}
then the above metric matches the large $\eta$ limit of (\ref{eq:hyp}), so indeed we are imposing we are approaching a hyperbolic black hole as we go towards the boundary. Moreover, the temperature of this hyperbolic black hole is 
\begin{equation}
    T_{\mathbb{H}_h} = \frac{(1-\rho_h)H(0)}{4\pi\rho_h r_c} = T_H,
\end{equation}
where $T_H$ is the temperature of the boundary event horizon, given in (\ref{eq:dS_temp}).

\subsubsection{The hyperbolic black hole $x=1$}
Of course, the bulk horizon meets the conformal boundary not only at the boundary event horizon, but also at the boundary cosmological horizon. Hence we can work similarly, to expand this point into another line in our integration domain by enforcing we approach another hyperbolic black hole horizon.

The work is very similar to the $x=0$ boundary. At $x=1$, we set the Dirichlet boundary conditions: $p_1=p_3=p_4=1$ and $p_2=p_5=0$. This time take
\begin{equation}
    \diff v = \inv{r_c}\diff \hat{t} - \frac{\rho_h}{H(1)(1-\rho_h)(1-y^2)}\diff{y}, \quad x = 1-H(1)\xi,
\end{equation}
then at the boundary $x=1$, which has become the surface $\xi=0$, the metric, to leading order in $\xi$ for each term, becomes
\begin{equation}
    \diff s^2\Big|_{\xi=0} = \frac{L^2}{y^2}\left[-\frac{(1-y^2)(1-\rho_h)^2H(1)^2}{\rho_h^2 r_c^2}\diff \hat{t}^2 + \frac{\diff y^2}{1-y^2} + \inv{4\xi^2}\diff \xi^2+\inv{4\xi}\diff\Omega^2_{(2)} \right],
\end{equation}
hence, just as above, the geometry is that of the limit of a hyperbolic black hole, now with temperature
\begin{equation}
    T_{\mathbb{H}_c} = \frac{(1-\rho_h)H(1)}{4\pi\rho_h r_c} = T_c.
\end{equation}

\subsubsection{The null hypersurface $y=1$}
We have one final boundary of our integration domain at $y=1$. Note that, the \textit{Ansatz} gives an inverse metric with a factor of $(1-y^2)$ in the $g^{yy}$ component, and hence the $y=1$ is a null hypersurface.

Recall that we solve only a subset of the components Einstein equation in the bulk, and we must solve the equations arising from the other components at some slice. So long as we do that, they will be satisfied everywhere. Hence, we set three Robin boundary conditions at $y=1$ by requiring that,
\begin{equation}
    E_{vv}=E_{vy}=E_{vx}=0.
\end{equation}
These three boundary conditions are sufficient to solve the PDEs.

Note that, \textit{a priori}, we cannot be sure whether this null hypersurface is the horizon of the bulk black hole. At the beginning of section \ref{sec:results} we'll provide numerical evidence that it is indeed the horizon by checking that there are future-directed, radial, null curves from any point with $y<1$ to the conformal boundary at $y=0$.

\subsection{Extracting the holographic stress tensor}\label{sec:ham_stress}

Once again, we can expand the functions off the boundary by ensuring order-by-order they solve the equations of motion. It can be shown that in Bondi gauge, no non-analytic terms will arise in such an expansion \cite{Fefferman:2007rka}. This lack of non-analytic terms ensures that the numerical method has exponential convergence even when reading asymptotic charges. This is in stark contrast with the DeTurck method which is typically plagued by non-analytic terms close to the AdS boundary (such as those in Eq.~(\ref{eq:drop_expansion})). We obtain
 \begin{equation}
     p_i(x,y) = \sum_{j=0}^5 \psi_i^{(j)}(x)y^j + \hdots\:, \quad \text{for}\; i=1,\hdots,4,
 \end{equation}
 where $\psi_1^{(0)}=\psi_3^{(0)}=\psi_4^{(0)}=1$ and $\psi_2^{(0)}=0$. The only functions not fixed by the local analysis of the equations of motion are $\{\psi_1^{(4)},\psi_2^{(4)},\psi_4^{(4)}\}$, so as expected we have three free functions at the conformal boundary. Moreover $\psi_2^{(4)}$ is fixed by the local analysis up to a constant:
 \begin{equation}\label{eq:C1}
     \psi_2^{(4)} =C_1 \frac{(1-x)x }{\left(1+2\rho_h-x\left(1-\rho_h^2\right)\right)^2}.
 \end{equation}
 We'll see that the value of this constant, $C_1$, plays a key role when we consider the flow along the horizon of the hammock. Next we go into Fefferman-Graham coordinates, so that the metric is in the form given by (\ref{eq:FG}). To do so, we take a transformation
 \begin{align}
     x &= w + \sum_{k=1}^5 \zeta_j(w) z^j \nonumber \\
     y &= \sum_{k=1}^5 \eta_j(w) z^j \nonumber \\
     v &= \frac{t}{2 r_c} + \sum_{k=1}^5 \theta_j(w) z^j,
 \end{align}
 where we match order-by-order to (\ref{eq:FG}), this time still taking $\Bar{g}_{\mu\nu}$ to be the de Sitter-Schwarzschild metric, but now in the coordinates given by (\ref{eq:dS-Schw_w}). 
 However, if we take the holographic stress tensor in these coordinates, we'll find it's not regular at the horizons, due to the flow along the bulk horizon. To alleviate this, we can take coordinates which are regular at \textit{both} the future event and future cosmological horizons in the boundary spacetime. That is, with the de Sitter-Schwarzschild we define a new time coordinate, $V$, by
 \begin{equation}\label{eq:ham_ingoing_coords}
     \diff t = \diff V - \frac{\rho_h(1-2w)}{(1-\rho_h)H(w)(1-w)w}\diff w.
 \end{equation}
 
 \begin{figure}[tb]
    \centering
    \begin{tikzpicture}
            \draw (0,4) -- (4,0) -- (0,-4) -- (-4,0) -- cycle;
            
            \draw [dashed] plot [smooth, tension=0.3] coordinates {(-3.7,0.3) (0,-3) (3.7,0.3)};
            \draw [dashed] plot [smooth, tension=0.7] coordinates {(-3.,1) (0,-1.2) (3,1)};
            \draw [dashed] plot [smooth, tension=0.9] coordinates {(-2.2,1.8) (0,0.5) (2.2,1.8)};
            \draw [dashed] plot [smooth, tension=1.2] coordinates {(-1.2,2.8) (0,2.2) (1.2,2.8)};
            
            \draw[above] (-2.5,1.7) node {$\mathcal{H}_h^+$};
            \draw[above] (2.5,1.7) node {$\mathcal{H}_c^+$};
            \draw[below] (-2.5,-1.7) node {$\mathcal{H}_h^-$};
            \draw[below] (2.5,-1.7) node {$\mathcal{H}_c^-$};
            
        \end{tikzpicture}
    \caption{The Penrose diagram of the four-dimensional de Sitter-Schwarzschild spacetime. The dashed curves are spacelike hyperslices of constant $V$, where $V$ is the coordinate defined in (\ref{eq:ham_ingoing_coords}). Hence the coordinate $V$ is regular at both the future event horizon, $\mathcal{H}_h^+$, and the future cosmological horizon, $\mathcal{H}_c^+$.}
    \label{fig:dS-S_Penrose}
\end{figure}
This brings the de Sitter-Schawrzschild metric to the form
 \begin{align}\label{eq:dS-S_V}
     \diff s^2 = \frac{r_h^2}{\left(1-w(1-\rho_h)\right)^2}\Bigg[-w(1-w) \frac{\left(1-\rho_h\right)^2}{\rho_h^2 r_c^2}&H(w)\diff V^2 - \frac{2(1-2w)(1-\rho_h)}{r_c \rho_h} \diff V \diff w 
     \nonumber \\
     & +\frac{4(1+\rho_h+\rho_h)^2}{1+2\rho_h-w(1-\rho_h^2)} \diff w^2 + \diff\Omega^2_{(2)}\Bigg].
 \end{align}
 In Figure \ref{fig:dS-S_Penrose}, we draw the Penrose diagram for the de Sitter-Schwarzschild geometry. The dashed curves are hypersurfaces of constant $V$.
 
 Now we can finally evaluate the holographic stress tensor via (\ref{eq:drop_stress}). We find that the stress tensor is regular everywhere, is conserved and has fixed trace, once again given by (\ref{eq:stress_trace}).
 
\subsection{Properties of the hammock horizon}\label{sec:ham_horizon}
Due to the fact that the bulk horizon of the black hammock is not Killing it can have further interesting properties, for example, it can have non-trivial expansion and shear. This has been observed in other flowing geometries such as those in \cite{Figueras:2012rb,Fischetti:2012vt,Santos:2020kmq}. In order to investigate these properties, one must consider the geometry of the bulk horizon, $\mathcal{H}^+$. From the \textit{Ansatz}, we know that the $y=1$ slice is a null hypersurface, but we cannot be immediately sure that it really is the horizon of the hammock. In the next section, we'll provide evidence that indeed the horizon is the $y=1$ slice of the bulk spacetime, so let us for now assume that as a given.

In general calculating the expansion and shear can be difficult, however, we follow \cite{Fischetti:2012vt,Santos:2020kmq}, which introduced a number of tricks that allows one to calculate the affine parameter along a generator of the horizon and from this quickly calculate the expansion and shear.

We assume the horizon is the $y=1$ hypersurface of the bulk spacetime, thus it is parameterised by $\{v,x,\theta,\phi\}$, where $\theta$ and $\phi$ are the usual angular coordinates in the two-sphere. Hence, $\mathcal{H}^+$ is a four-dimensional null hypersurface with a three-dimensional space of generators. Due to stationarity and spherical symmetry, the spacetime also possesses three Killing vector fields, which we will denote as $\partial_I$, for $I=v,\,\theta,\,\phi$, none of which are generators of the horizon, which is non-Killing. Thus, any two generators of the horizon will be related by the action of a combination of the $\partial_I$ vector fields. This means for a given value of $x$, we can arbitrarily pick a horizon generator, say with affine parameter, $\lambda$. We can extend $\lambda$ to a function of $x$ along the horizon by requiring that it is independent of $v,\,\theta$ and $\phi$, so that $\lambda=\lambda(x)$.

Let $U^a$ be the tangent vector to the horizon generator with affine parameter $\lambda$. We know that each of the $\partial_I$ vectors is tangent to $\mathcal{H}^+$, hence $U \perp \partial_I$. Now, one can define another vector field $S^a$ such that $S^a U_a = -1$, $S \perp \partial_I$ and $S^a S_a = 0$. 

In order to find the expansion and shear, first one defines the tensor $B_{ab}=\nabla_b U_a$, which is symmetric since $U^a$ is hypersurface orthogonal. Now let us consider a deviation vector, $\eta$, orthogonal to both $U$ and $S$. Then
\begin{equation}
    \eta^bB^a{}_b=U^b\nabla_b\eta^a,
\end{equation}
that is, $B^a{}_b$ measures the failure of a deviation vector to be parallely transported along $U$. Since $U$ and $\partial_I$ commute, $\partial_I$ are deviation vectors for the geodesic congruence, so from the above equation,
\begin{equation}
    \left(\partial_I\right)^bB^a{}_b=U^b\nabla_b\left(\partial_I\right)^a.
\end{equation}

Now let us consider $h_{IJ}=\partial_I \cdot \partial_J$. In the $\{v,x,y,\theta,\phi\}$ coordinates, $h_{IJ}=g_{IJ}$, so it is simply the induced metric on a constant $x$ and constant $y$ slice. If one differentiates the component of $h_{IJ}$ with respect to the affine parameter $\lambda$, one finds that
\begin{align}
    \der{\lambda} h_{IJ} &= U^b\nabla_b\left(\partial_I \cdot \partial_J\right) \nonumber \\
    &= \left(\partial_I\right)^b B^a{}_b\left(\partial_J\right)_a + \left(\partial_J\right)^b B^a{}_b\left(\partial_I\right)_a \nonumber \\
    &= 2 B_{ab}\left(\partial_I\right)^a \left(\partial_J\right)^b \nonumber\\
    &= 2 B_{IJ}.
\end{align}

Now, when calculating the expansion, shear and twist of a null geodesic congruence, it turns out the space one has to work with is an equivalence class of deviation vectors, where first we restrict to vectors which are orthogonal to $U^a$, and then consider two such deviation vectors as equivalent if they differ by a multiple of $U^a$ (see, for example, Section 9.2 of \cite{Wald:1984rg} for further details). Let us denote the vector space of such an equivalence class as $\hat{V}$. This is a $3=5-2$ dimensional vector space in our case, where we have five bulk dimensions. A vector/covector in the spacetime naturally gives rise to a vector/covector in $\hat{V}$ if and only if its contraction with $U$ is zero. Furthermore, a general tensor $T^{a_1 \hdots a_k}{}_{b_1\hdots b_\ell}$ in the spacetime naturally gives rise to a tensor $\hat{T}^{a_1\hdots a_k}{}_{b_1\hdots b_\ell}$ if and only if contracting any one of its upper or lower indices with $U_a$ or $U^a$ and then the remainder of indices with vectors or covectors with natural realisations in $\hat{V}$ gives zero.

Each of $g_{ab},\, B_{ab}$ and $\left(\partial_I\right)^a$ satisfy the above definitions, so they naturally give rise to tensors in $\hat{V}$. Therefore, the three linearly independent vectors $\left(\partial_I\right)^a$ provide a basis for $\hat V$, so for any spacetime tensor $T$ satisfying the above condition, we can find $\hat T$ simply by reading off the $\{v,\theta,\phi\}$ components of $T$. That is,
\begin{align}
    \hat{h}_{IJ} &= h_{IJ} = g_{IJ}, \\
    \hat{B}_{IJ} &= B_{IJ} = \inv{2}\der{\lambda} h_{IJ} \label{eq:hatB}.
\end{align}
Now that we're working in the vector space $\hat{V}$, we can define the \textit{expansion} and \textit{shear}, respectively, as
\begin{align}
    \Theta &= \hat{B}^I{}_I = \hat{h}_{IJ}\hat{B}^{IJ} = h_{IJ}B^{IJ}, \label{eq:expansion} \\
    \sigma_{IJ} &= \hat{B}_{IJ} - \inv{d-2}\Theta  \hat{h}_{IJ} = B_{IJ} - \inv{d-2}\Theta  h_{IJ}, \label{eq:shear}
\end{align}
where $d$ is the dimension of the bulk spacetime, so that, in our case $d=5$. The \textit{twist} is the anti-symmetric part of $\hat{B}_{IJ}$, which vanishes here, due to the symmetry of $B_{IJ}$.

We already have $h$, so we need only calculate $B$ to find the expansion and shear. This can be done by finding the affine parameter $\lambda(x)$, and then solving (\ref{eq:hatB}) for $B$. In order to find the affine parameter, we can utilise Raychaudhuri's equation, which is given by
\begin{equation}
    \der[\Theta]{\lambda} = - \hat{B}^b{}_a \hat{B}^a{}_b - R_{ab}U^a U^b.
\end{equation}
By the Einstein equation, $R_{ab} \propto g_{ab}$, and moreover $U^a$ is null, so that the final term above vanishes. Also, $\hat{B}^{ab}$ is symmetric and orthogonal to both $U$ and $S$, so the above equation simplifies to
\begin{equation}
    \der[\Theta]{\lambda} = -\hat B^{IJ} \hat B_{IJ}.
\end{equation}
One can then substitute $\Theta$ and $\hat{B}_{IJ}$ in the above equation for expressions in terms of $h_{IJ}$ by using (\ref{eq:hatB}) and (\ref{eq:expansion}), which yields the following second order ODE for $\lambda$:
\begin{equation}\label{eq:lambda}
    \lambda'' = \lambda'\left(h^{IJ}h'_{IJ}\right)^{-1}\left[\inv{2}h^{IJ}h^{KL}h'_{IK}h'_{JL} + \left(h^{IJ}h'_{IJ}\right)'\right],
\end{equation}
where $'$ denotes a derivative with respect to $x$, the coordinate along the horizon. A more general ODE is given in \cite{Santos:2020kmq} in the case where the horizon is not necessarily a constant $y$ slice, but rather some non-trivial surface in the $(x,y)$ plane.

Once one has a numerical solution for the line element of the black hammock, one can numerically solve the above ODE for the affine parameter $\lambda(x)$. Finally with such an affine parameter, one can evaluate $\hat{B}_{IJ}$ since (\ref{eq:hatB}) implies that
\begin{equation}\label{eq:hatB2}
    \hat{B}_{IJ} = \inv{2}\left(\lambda'(x)\right)^{-1}\der{x}h_{IJ}.
\end{equation}
From this the expansion and shear along the horizon can easily be computed using (\ref{eq:expansion}) and (\ref{eq:shear}).

\section{Results}\label{sec:results}
\paragraph{The position of the hammock horizon.}
The \textit{Ansatz} for the black hammock fixed that the $y=1$ slice was a null hypersurface in the bulk spacetime, however, at no point in the \textit{Ansatz} or in the boundary conditions did we explicitly set that it really was the bulk horizon. In particular, we cannot be sure it is not instead some inner horizon, and that the real event horizon lies outside of it. We can check this isn't the case by ensuring we can reach the conformal boundary via a future-directed, causal curve starting at any given point outside the $y=1$ hypersurface.

Let us consider, in $(v,x,y,\theta,\phi)$ coordinates, a radial curve with tangent vector given by
\begin{equation}
    U^a = \left(\inv{x(1-x)p_2(x,y)}U^x,\,U^x,\,U^y ,\,0,\,0\right).
\end{equation}
The condition for such a curve to be null leads to the equation
\begin{equation}
    \der[y]{x} = -\frac{(1-\rho_h)(1-y^2)p_1(x,y)H(x)}{2\rho_h(1-x)x\, p_2(x,y)}.
\end{equation}
We numerically solve the above ODE for $y(x)$. This provides a family of null curves, parameterised by the choice of initial condition of the ODE. The $y$ coordinate depends monotonically on the $x$ coordinate, so we can indeed use $x$ as a parameter along the curve. One finds that if one takes as an initial condition that $y=1$ for some value of $x$, then one finds that the curve remains on the $y=1$ hypersurface for all time. On the other hand, if one takes $y=1-\epsilon$ as an initial condition for any $\epsilon > 0$, then the curve will always intersect the conformal boundary at $y=0$. Moreover, we can easily see that the vector $U^a$ is future-directed, since its inner product with $\partial / \partial v$ is negative in the asymptotic region where $\partial / \partial v$ defines the time-orientation. Thus we can find a future-directed causal curve from any point in the exterior of $y=1$ to the conformal boundary at $y=0$. This proves that $y=1$ is not an inner horizon, and provides strong evidence that the $y=1$ hypersurface is indeed the horizon of the black hammock.

\paragraph{Embeddings of the horizons.}
In order to aid with visualisation of the geometries, we can embed a spatial cross-section of the horizons into hyperbolic space. The cross-section of the horizon of the black hammock and each of the horizons of the black tunnel are three-dimensional, so we can embed them into four-dimensional hyperbolic space $\mathbb{H}^4$, which has metric
\begin{equation}
    \diff s^2_{\mathbb{H}^4} = \frac{L_4^2}{Z^2}\left(\diff Z^2 + \diff R^2 +R^2\diff \Omega_{(2)}^2 \right).
\end{equation}
Let us consider finding an embedding of the horizons of the black tunnel. The bulk event and bulk cosmological horizons are, respectively, the $x=0$ and $x=1$ surfaces, and are parameterised by the $y$ coordinate. Hence, for each of the horizons we seek an embedding of the form $(R(y),Z(y))$. Such a surface in $\mathbb{H}_4$ has induced metric
\begin{equation}
    \diff s^2_{emb} = \frac{L_4^2}{Z(y)^2}\left(\left(Z'(y)^2 + R'(y)^2 \right)\diff y^2 +R(y)^2\diff \Omega_{(2)}^2 \right).
\end{equation}
We compare the above line element with the induced metric of a constant time slice of each of the bulk event and bulk cosmological horizons. In each case, this yields a simple first-order ODE and an algebraic equation which can be solved numerically for $R(y)$ and $Z(y)$. Note that the embeddings of the two horizons (the bulk event horizon at $x=0$ and the bulk cosmological horizon at $x=1$) must be found separately, but can be embedded into the same space. There is a fair amount of freedom in the embeddings corresponding to the initial conditions of the ODEs, so really it is only the shape of the horizons that matters. We choose $L_4 = L$, and take for the cosmological horizon the initial condition that $R(0)=1$, and for the event horizon we pick the initial condition that $R(0)=\rho_h$. 

\begin{figure}[tb]
    \centering
     \begin{subfigure}{0.49\textwidth}
         \includegraphics[width=\linewidth]{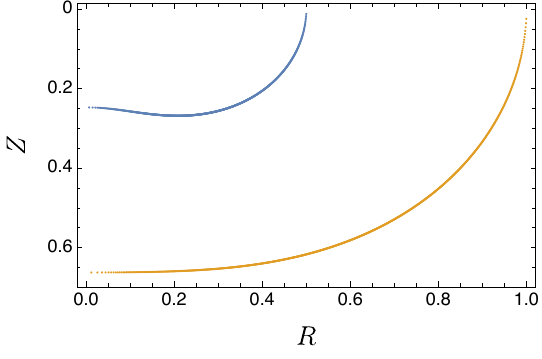}
        \caption{The embedding of the black tunnel}
        \label{fig:drop_emb}
     \end{subfigure}
     \centering
     \begin{subfigure}{0.49\textwidth}
        \includegraphics[width=\linewidth]{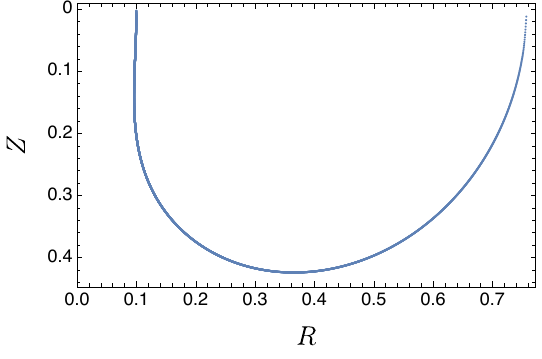}
        \caption{The embedding of the black hammock}
        \label{fig:ham_emb}
     \end{subfigure}
     \caption{The embedding diagrams of the black tunnel and hammock with $\rho_h = 0.5$ in four-dimensional hyperbolic space. The tunnel has two disconnected horizons; the blue curve is the bulk event horizon whilst the orange curve is the bulk cosmological horizon. The hammock possesses a single connected horizon.}\label{fig:embeddings}
\end{figure}

To obtain the embedding diagram, one then plots the resultant values of $\{R(y),Z(y)\}$ across the range $y\in(0,1)$ for each of the horizons. In figure \ref{fig:drop_emb}, we've plotted together the embedding diagrams for the two horizons of the black tunnel, with a value of the parameter $\rho_h=0.5$.


We can apply a similar procedure for the black hammocks, which only has one horizon, which in the coordinates of our \textit{Ansatz} is the $y=1$ hypersurface, hence this time is parameterised by the $x$ coordinate. Once again we compare the line element of the embedding $(R(x),Z(x))$ in $\mathbb{H}_4$ to the induced metric on a constant time slice of the bulk horizon of the hammock. This yields an ODE, which we can solve with the initial condition $R(0)=0.1$. The embedding diagram obtained for the black hammock is plotted in figure \ref{fig:ham_emb}.


\paragraph{The ergoregion of the hammocks.}
By above, we know that the horizon of the black hammock is the $y=1$ null hypersurface. However, note that the Killing vector field $k^a = \left(\partial / \partial v\right)^a$ may not be timelike for the whole of the exterior region. The region for which $k^a$ is spacelike is called the \textit{ergoregion}, which is bounded by the horizon and the surface at which $k^2 = 0$, called the \textit{ergosurface}. 

\begin{figure}[b!]
    \centering
    \includegraphics[width=0.8\linewidth]{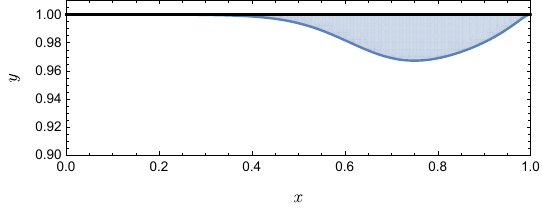}
    \caption{The ergoregion of the black hammock, with $\rho_h=0.1$, plotted as in the $(x,y)$ coordinate space. The black line at $y=1$ is the horizon of the black hammock. The blue curve is the ergosurface where the Killing vector $k = \partial / \partial v$ is null and the shaded region is the ergoregion where $k$ is spacelike.}
    \label{fig:ergo}
\end{figure}

We find that for large values of the radius ratio, $\rho_h$, the ergosurface is very close to the horizon, meaning the ergoregion is extremely small. For smaller values of $\rho_h$, however, the ergoregion is noticeably larger. In Figure \ref{fig:ergo} we have plotted the ergoregion as a surface in the $(x,y)$ plane for a black hammock with $\rho_h=0.1$. As one would expect from the \textit{Ansatz}, the ergosurface approaches the horizon as one goes towards $x=0$ or $x=1$, which we recall are the points on the conformal boundary at which the bulk horizon approaches the geometry of a hyperbolic black hole.

In general, the presence of an ergoregion can lead to superradiant instabilities of black holes \cite{Green:2015kur}. Noting the fact that as the radius ratio, $\rho_h$, is taken to be small (\textit{i.e.} as the horizons of the de Sitter-Schwarzschild boundary geometry are taken to be far apart) the ergoregion becomes larger and larger, it would be particularly interesting to investigate whether the black hammocks for these values of $\rho_h$ do suffer such superradiant instabilities.

\paragraph{Energy.}
Now we'll turn to some of the conserved charges that the solutions possess. Firstly we'll look at the energy of both the tunnels and the hammocks, and then the flow which is non-zero only for the hammocks. Before we do so we would like to stress one particularly interesting point regarding the normalisation of these quantities. The energy and flow are dimensionful quantities, so in order to compare the quantities between solutions, one has to multiply by a suitable power of a dimensionful parameter of the solutions. Often the temperature of the black hole is used as this normalising factor. However, in our case, we have two temperatures which are not equal: the temperature of the event horizon, $T_H$, and the temperature of the cosmological horizon, $T_c$. This is connected to the fact that in de Sitter-Schwarzschild, one has two scales, the mass of the black hole and the de Sitter length scale. It is not immediately clear which of the temperatures, $T_H$ or $T_c$, is the meaningful quantity to normalise with, or if instead we should use some combination of them. Indeed, the fact that the two temperatures of the horizons are different means that it is difficult to define a canonical ensemble, and hence it is hard to discuss the solutions using thermodynamic analysis. We'll return to this point in the discussion in section \ref{sec:discussion}. For now, let us present the charges which we choose to make dimensionless using $T_H$, which seems to yield results without divergent behaviour at the endpoints of the parameter space.

The energy of the black tunnels and hammocks can be defined from the holographic stress tensor, $T_{\mu\nu}$, in the standard way \cite{Henningson:1998gx, Balasubramanian:1999re}. One takes a Cauchy slice $\Sigma$ of the boundary de Sitter-Schwarzschild geometry and then the energy is defined by an integral over this slice as
\begin{equation}
    E := -\int_{\Sigma}\diff^3 x \sqrt{h}n^\mu k^\nu T_{\mu\nu},
\end{equation}
where $k^\mu$ is the stationary Killing vector field of the de Sitter-Schwarzschild metric and $h$ the determinant of the induced metric on $\Sigma$ and $n^\mu$ the unit normal to $\Sigma$.



For the hammocks, we have to be careful to use coordinates on which the stress tensor is regular at both horizons, which can be done by taking the de Sitter-Schwarzschild metric in the form given in (\ref{eq:dS-S_V}).

\begin{figure}[tb]
    \centering
     \begin{subfigure}{0.49\textwidth}
         \includegraphics[width=\textwidth]{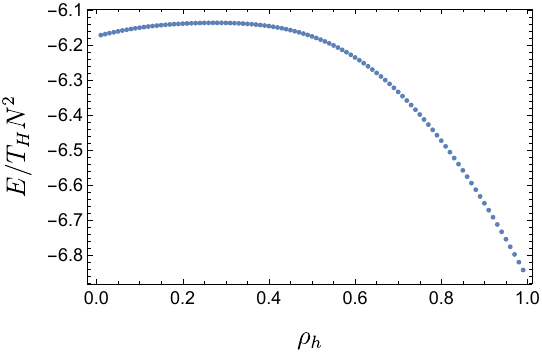}
         \caption{Energy of the black tunnels}
         \label{fig:drop_ETh}
     \end{subfigure}
     \centering
     \begin{subfigure}{0.49\textwidth}
         \includegraphics[width=\textwidth]{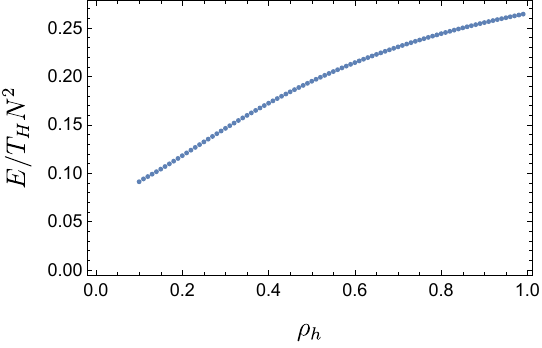}
         \caption{Energy of the black hammocks}
         \label{fig:ham_ETh}
     \end{subfigure}
     \caption{The plots of the energy of the black tunnel and black hammock solutions across different values of the parameter $\rho_h$. Here we have divided the energy, $E$, by the temperature of the event horizon, $T_H$, in order to obtain a dimensionless quantity.}\label{fig:energy}
\end{figure}

We pick the surface $\Sigma$ to be a constant $V$ slice, and so $n_\mu \propto (\diff V)_\mu$. Such surfaces are shown as dashed curves in the Penrose diagram for de Sitter-Schwarzscild, which is shown in Figure \ref{fig:dS-S_Penrose}. In these coordinates, the stationary Killing vector field is given by $k^\mu = (\partial / \partial V)^\mu$. With these choices, the integral for the energy becomes
\begin{equation}
    E = -2\pi\,r_c^3\,\rho_h^3\,(1-\rho_h)\int_0^1 \diff w \frac{T^V{}_V}{\left(1-w(1-\rho_h)\right)^4}.
\end{equation}
The integral can be computed numerically for both the tunnels and the hammocks using the holographic stress tensors found in sections \ref{sec:drop_stress} and \ref{sec:ham_stress}, respectively. It is immediately clear that the stress tensor for the tunnels is regular and finite at the horizons, so there are no difficulties in calculating the energy for each of the values of the radius ratio, $\rho_h$. We've plotted the energy of each tunnel solution in Figure \ref{fig:drop_ETh}. One particularly interesting and surprising feature of this plot is that the energy of the tunnels (when normalised by the event horizon temperature) is a non-monotonic function of the parameter $\rho_h$. However, the fact that one finds a non-monotonic function, and moreover the position of the maximum, is dependent on the fact we've used $T_H$ to make the energy dimensionless.

The regularity of the stress tensor of the black hammocks depends on non-trivial relations between (fourth) derivatives of the metric functions. It is expected that the true, full solutions will satisfy these relations, however there is some noise in the numerical solutions due to the fact the solutions have been found on a discrete grid. Thus, in order to ensure these relations are satisfied when finding the metric functions numerically, one has little choice other than to use high precision and a large number of lattice points in the discretization (we used a precision of 50 decimal places on the $120 \times 120$ grid). We found that the smaller the value of $\rho_h$, the more difficult it is to ensure that one attains a regular stress tensor numerically. We were able to get sensible results for $\rho_h \geq 0.1$, and have plotted the energy for these solutions in Figure \ref{fig:ham_ETh}. We did find hammock solutions for smaller values of $\rho_h$, but not at a high enough resolution to ensure we obtained a regular stress tensor numerically and hence we've omitted these points from the plot.

\paragraph{Flow.}
Now let us turn our attention to some additional properties of the hammocks. Unlike the tunnel, the hammock is not a static solution in the bulk. This is due to the fact that the horizon is not a Killing horizon; there is flow along it. This flow is dual to the fact that the CFT is deconfined in the hammock phase. This flow can be defined in terms of the holographic stress tensor by an integral over a two-sphere of fixed radius $r$ as follows:
\begin{equation}
    \Phi := -\int_{S_r^2} \diff^2 x \sqrt{-\gamma}\,m^\mu k^\nu T_{\mu\nu},
\end{equation}
where $\gamma_{\mu\nu}$ is the induced metric on a constant radius slice of the boundary de Sitter-Schwarzschild geometry with determinant $\gamma$ and unit normal $m^\mu$, whilst $k^\mu$ is again the stationary Killing vector field. Due to the conservation of the stress tensor, this integral is invariant of the choice of the radius of the two-sphere one integrates over.

Once again, we work in the coordinates $(V,w,\theta,\phi)$ of (\ref{eq:dS-S_V}) in which the stress tensor is regular at the horizons. Thus, we consider a constant $w$ slice, with $m_\mu \propto (\diff w)_\mu$ and $k^\mu = (\partial / \partial V)^\mu$. This integral can actually be evaluated up to a constant $C_1$ directly from the expansion of the equations of motion about the conformal boundary so that the flow is given by
\begin{equation}
    \Phi = \frac{N^2(1-\rho_h)}{32 \pi\,r_c^2\,\rho_h\left(1+\rho_h+
    \rho_h^2\right)^2} C_1,
\end{equation}
where $C_1$ can be obtained numerically from a fourth derivative of the metric function $p_2(x,y)$ using the equation (\ref{eq:C1}). As expected this result is independent of the radial coordinate $w$.

This calculation confirms that in the deconfined hammock phase, the CFT on the field theory side of the duality flows at order $\bigO{N^2}$.

\begin{figure}[tb]
    \centering
    \includegraphics[width=0.6\linewidth]{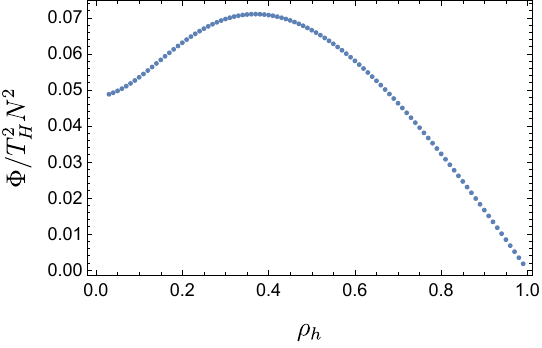}
    \caption{The flow along the black hammock, divided by the temperature of the event horizon squared, for different values of the radius ratio, $\rho_h$. The flow, $\Phi$, is a non-monotonic function of $\rho_h$.}
    \label{fig:flow}
\end{figure}

Since one only needs to extract a constant using the metric functions found numerically, the flow can be found to a high degree of accuracy for each of the black hammock solutions we obtained. In Figure \ref{fig:flow}, we have plotted the flow as a function of the radius ratio, $\rho_h$. As expected, $\Phi$ is positive for each of the solutions, which corresponds to flow from the hotter event horizon to the cooler cosmological horizon. Moreover, the flow tends towards to zero in the extremal limit ($\rho_h \to 1$) in which the event and cosmological horizons are aligned and have the same temperature. 

Interestingly, the flow is a non-monotonic function of $\rho_h$, with a maximum at $\rho_h \simeq 0.396$. This property was also found in the flowing black funnel solutions of \cite{Santos:2020kmq}, which are dual to a CFT on a flat Schwarzschild background with the CFT asymptotically in a thermal state with non-zero temperature. As of yet, there is no field theory explanation of this non-monotonic behaviour of the flow for either the flowing funnels or the hammocks. The position of the maximum of the flow for the black hammocks is at a different value of the ratio of the two temperatures than the flowing funnels, which is perhaps to be expected since in the case of the hammocks, the distance between the two horizons is finite and naturally varies as one changes the ratio of their temperatures, whereas the event horizon in the flowing funnels is always an infinite distance from the asymptotic region.

In \cite{Santos:2020kmq} it was conjectured, with evidence provided by local Penrose inequalities \cite{Figueras:2011he}, that the turning point of the flow is located at the same point in the parameter space as the phase transition between the confined and deconfined phase of the CFT. It would be of great interest to investigate whether this also holds in the case of the hammocks, with the turning point of the flow similarly signifying a transition from dominance of the tunnel solution to that of the hammock solution.

However, let us once again stress that the fact that one finds a non-monotonic function, and moreover the position of the maximum, depends on the quantity one uses to make the flow dimensionless. If one uses $T_c$ instead of $T_H$, one finds that the flow diverges as $\rho_h \to 0$. It is not obvious which quantity one should use.

\paragraph{Expansion and shear of the hammock horizon.} 
The fact that the bulk horizon of the hammock is non-Killing allows it to have interesting properties, for example, non-trivial expansion and shear.

\begin{figure}[tb]
    \centering
     \begin{subfigure}{0.335\textwidth}
         \includegraphics[width=\linewidth]{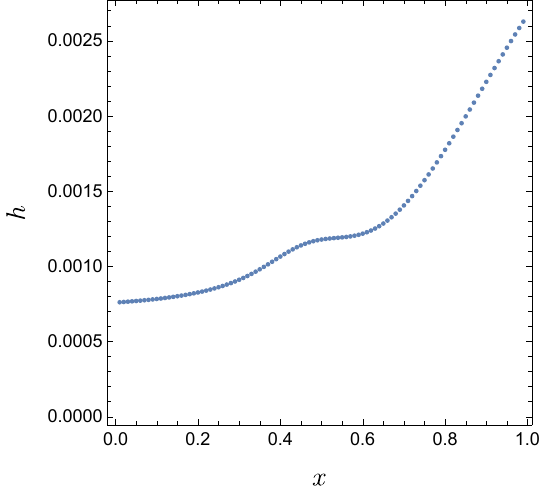}
        \caption{The determinant}
        \label{fig:det}
     \end{subfigure}
     \centering
     \begin{subfigure}{0.32\textwidth}
         \includegraphics[width=\linewidth]{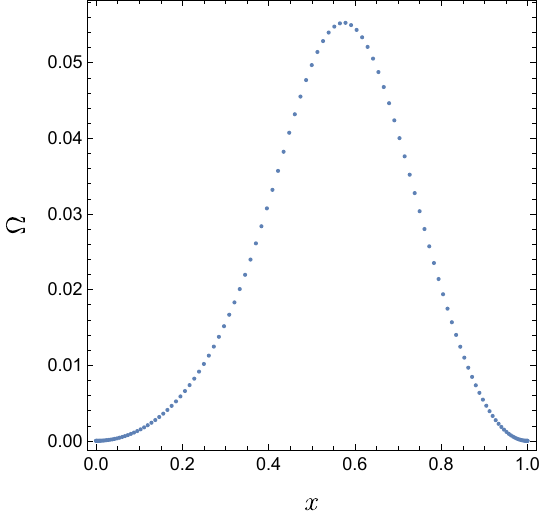}
         \caption{The coordinate velocity}
         \label{fig:vel}
     \end{subfigure}
     \begin{subfigure}{0.325\textwidth}
        \centering
        \includegraphics[width=\linewidth]{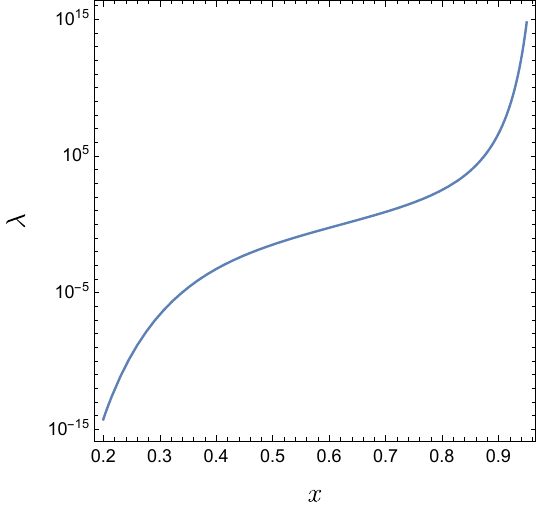}
        \caption{The affine parameter}
        \label{fig:affine}
     \end{subfigure}
     \caption{Here we plot the the determinant, $h = \det h_{IJ}$, the coordinate velocity $\Omega(x)$, and the affine parameter, $\lambda(x)$, in subfigures \textbf{(a)}, \textbf{(b)} and \textbf{(c)}, respectively, against the coordinate $x$ along the horizon for the black hammock with $\rho_h=0.4$. Clearly, the determinant is a monotonically increasing function of $x$. Meanwhile  the coordinate velocity $\Omega(x)$ is positive, showing that the flow will be from the boundary event horizon to the boundary cosmological horizon. The affine parameter is also monotonically increasing, from zero at $x=0$, at the boundary event horizon, up to positive infinity at $x=1$, at the boundary cosmological horizon.}
     \label{fig:det_and_vel}
\end{figure}

Firstly, let us consider the behaviour of $h_{IJ}$, which was defined in section \ref{sec:ham_horizon}, along the horizon, the $y=1$ hypersurface. In particular, we plot its determinant, $h = \det h_{IJ}$, in Figure \ref{fig:det}. Clearly, $h$ monotonically increases with $x$, as one moves along the horizon from the boundary event horizon (at $x=0$) to the boundary cosmological horizon (at $x=1$).

Moreover, from the line element, (\ref{eq:ham_Ansatz}), the coordinate velocity along the horizon is given by $\Omega(x) = x(1-x)p_2(x,1)$. We plot this in Figure \ref{fig:vel} against the $x$ coordinate. For each value of the radius ratio, $\rho_h$, the coordinate velocity is positive across the horizon, which supports the fact that the flow along the horizon is from the hotter boundary event horizon at $x=0$ to the cooler boundary cosmological horizon at $x=1$. This fact, together, with the fact that the determinant of $h_{IJ}$ is an increasing function of $x$, suggests strongly that the past horizon lies at $x=0$.


Now let us turn our attention to the affine parameter, the expansion and the shear. Following the discussion in section \ref{sec:ham_horizon}, in order to obtain the affine parameter, $\lambda(x)$, we can numerically solve the ODE given by (\ref{eq:lambda}). In Figure \ref{fig:affine}, we plot $\lambda(x)$ against $x$. As one would expect, $\lambda$ increases monotonically with $x$. It diverges as $x \to 1$ at the boundary cosmological horizon. At the past horizon, $x=0$, $\lambda$ goes instead to a finite value, which we are free to choose by rescaling the affine parameter. We have used such freedom to fix $\lambda(0)=0$ at the past horizon and $\lambda'(0.5)=1$.

\begin{figure}[tb]
    \centering
     \begin{subfigure}{0.425\textwidth}
         \includegraphics[width=\linewidth]{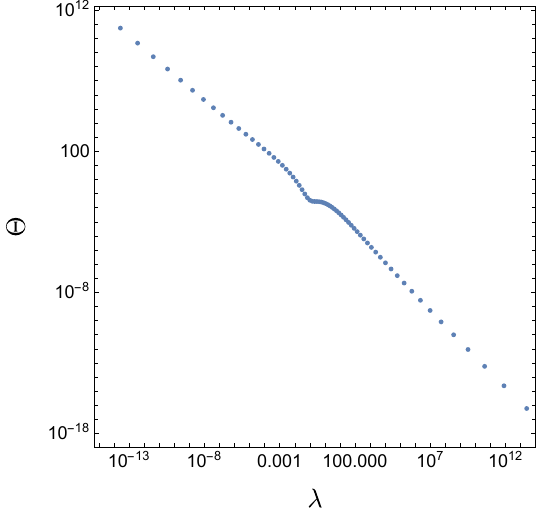}
     \end{subfigure}
     \centering
     \begin{subfigure}{0.42\textwidth}
         \includegraphics[width=\linewidth]{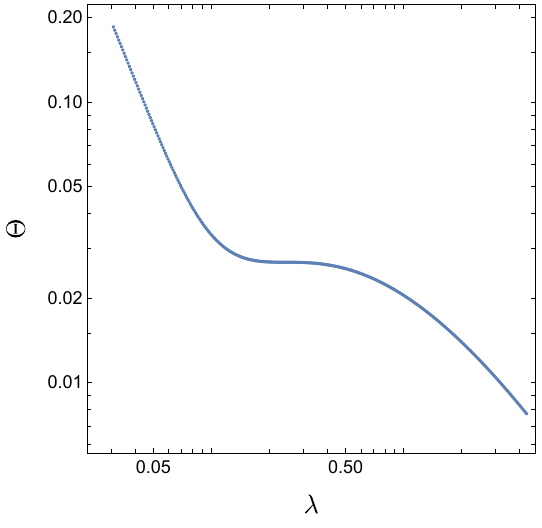}
     \end{subfigure}
     \caption{The expansion, $\Theta$, of the horizon plotted against the affine parameter, $\lambda$. The left-hand panel is a log-log plot of this curve across a large portion of the horizon. The right hand panel is a zoom (still with a logarithm scale of both axes) of the plot on the left, focusing on the portion of the curve with non-linear behaviour of the expansion.} 
     \label{fig:expansion}
\end{figure}

Once we have the affine parameter, the expansion and shear are easy to calculate using (\ref{eq:hatB2}) followed by (\ref{eq:expansion}) and (\ref{eq:shear}). In Figure \ref{fig:expansion}, we plot the expansion along the horizon (now parameterised by the affine parameter $\lambda$) in a log-log plot. As expected by Raychaudhuri's equation, we find that $\diff \Theta / \diff \lambda < 0$ everywhere. For small and large values of $\lambda$, the expansion follows a line in the log-log plot. However, there is some interesting non-trivial behaviour of the expansion, in this case at around $\lambda \simeq 0.2$. This behaviour is partially explained by the behaviour of the shear.

\begin{figure}[tb]
    \centering
     \begin{subfigure}{0.45\textwidth}
         \includegraphics[width=\linewidth]{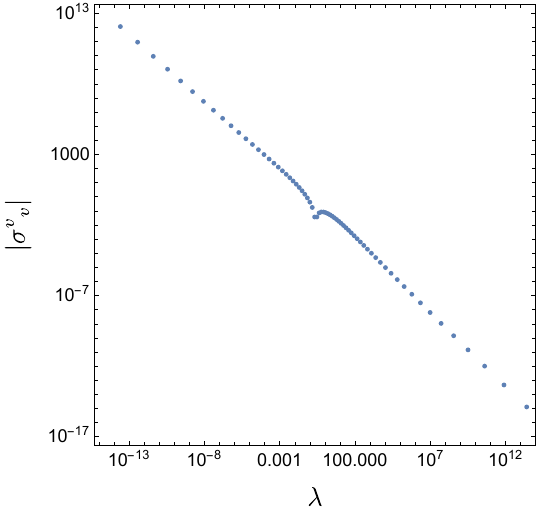}
     \end{subfigure}
     \centering
     \begin{subfigure}{0.45\textwidth}
         \includegraphics[width=\linewidth]{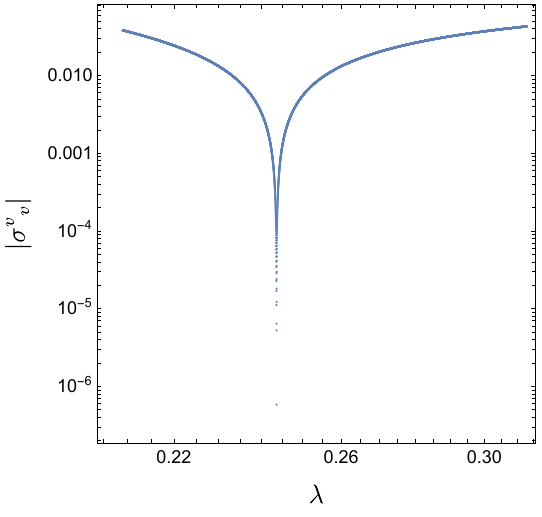}
     \end{subfigure}
     \vspace{0.5cm}\,
     \begin{subfigure}{0.45\textwidth}
         \includegraphics[width=\linewidth]{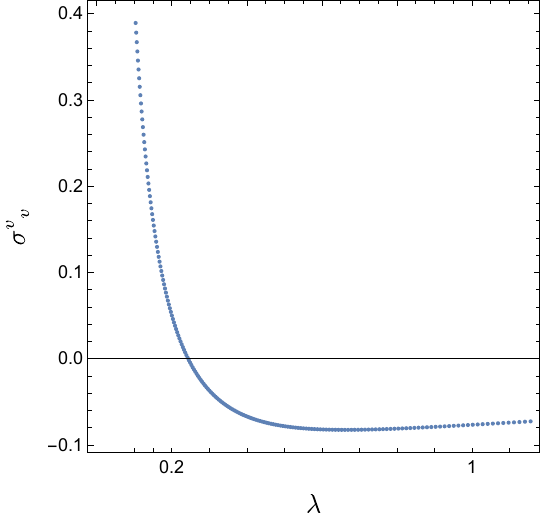}
     \end{subfigure}
     \caption{In the top-left panel, the absolute value of the $vv$ component of the shear, $\abs{\sigma^v{}_v}$, is plotted against the affine parameter, $\lambda$, along the horizon of the black hammock with $\rho_h=0.4$ in a log-log plot. We have simply taken the absolute value in order to be able to take the logarithm. The plot in the top-left panel indicates some interesting behaviour of $\abs{\sigma^v{}_v}$ near $\lambda \simeq 0.24$, which we have zoomed in on in the top-right panel, after interpolating between lattice points. The behaviour near this point is more easily understood by looking at the behaviour of $\sigma^v{}_v$ nearby, which is plotted (in a standard plot, without taking logs on either axis) in the bottom panel. The shear vanishes at $\lambda \simeq 0.2437$, with each of its non-trivial components, $\sigma^I{}_J$, flipping signs across this point on the horizon. We find $\sigma^v{}_v$ is positive below this value of $\lambda$ and negative for large values of $\lambda$, taking its minimum value at $\lambda \simeq 0.2738$, before tending towards zero as $\lambda$ diverges.} 
     \label{fig:shear}
\end{figure}

The shear in $\{v,\theta,\phi\}$ coordinates is given by a diagonal, traceless $3\times3$ matrix, with its $\theta$ and $\phi$ components related due to spherical symmetry. Hence, the shear at each point is completely determined by its $vv$ component:
\begin{equation}
    \sigma^I{}_J = \text{Diag}\left(\sigma^v{}_v, -\inv{2}\sigma^v{}_v,-\inv{2}\sigma^v{}_v\right).
\end{equation}
In Figure \ref{fig:shear}, we plot $\sigma^v{}_v$ against the affine parameter, after taking the absolute value so that we can plot in a log-log plot. For small values of $\lambda$ (nearer to the boundary event horizon) the $\sigma^v{}_v$ component of the shear is positive, whereas, for large values of $\lambda$ (nearer the boundary cosmological horizon), $\sigma^v{}_v$ is negative.
By continuity, the shear must vanish at some point along the horizon. This leads to the ``kink'' in the curve of the absolute value of $\sigma^v{}_v$, seen in Figure \ref{fig:shear} located at around $\lambda \simeq 0.2437$.
In the top-right panel of Figure \ref{fig:shear}, we've zoomed in on this point further, using lots of interpolation between the lattice points. However, this apparent ``kink'' is only an artifact of the fact we've taken the absolute value so that we can plot in a log-log plot. To explain the behaviour more clearly, we've plotted $\sigma^v{}_v$ in the region of interest without taking the absolute value or the log on either axis in the bottom panel of Figure \ref{fig:shear}. Indeed, the shear vanishes at $\lambda \simeq 0.2437$, and $\sigma^v{}_v$ has a minimum value at $\lambda \simeq 0.2738$ before returning to zero as $\lambda$ becomes large. This behaviour of the shear vanishing at a single point along the horizon, and flipping signs across it seems to occur for each of the hammocks we found across the parameter space.

As discussed above, the expansion also seems to have interesting behaviour near this value of the affine parameter, which we've focused in on on the right hand panel of Figure \ref{fig:expansion}. This behaviour follows from the behaviour of the shear, since by Raychaudhuri's equation,
\begin{equation}
    \der[\Theta]{\lambda} = -\inv{3}\Theta^2 - \sigma^{IJ}\sigma_{IJ},
\end{equation}
so that when the shear vanishes, the magnitude of $\diff\Theta/\diff\lambda$ decreases, so that the plot of $\Theta$ temporarily flattens out near this point (though the gradient never vanishes completely). Indeed, we checked explicitly that the values found for the expansion and shear accurately satisfy Raychaudhuri's equation.




\section {Discussion}\label{sec:discussion}
We constructed the holographic duals to a large $N$, strongly coupled CFT living on a de Sitter-Schwarzschild background. That is, we found the five-dimensional, AlAdS spacetimes with a de Sitter-Schwarzschild geometry on the conformal boundary. There are two solutions: the black tunnel solution with two disconnected bulk horizons, which is dual to a confined phase of the CFT, and the black hammock solution with a single connected bulk horizon, which is dual to a deconfined phase of the CFT. We were able to find black tunnel solutions across the whole of the parameter space, $\rho_h \in (0,1)$, where the radius ratio, $\rho_h$, is the ratio of the proper radii of the boundary event and cosmological horizons, and we were able to find black hammock solutions for a large range of the parameter space ($\rho_h \geq 0.03$).

We used complementary methods to obtain the two families of solutions. In order to find the black tunnels, we used the DeTurck method, which is very effective for static problems, yielding elliptic PDEs. Moreover with the boundary conditions in our case, it can be proven that Ricci solitons cannot exist, so that any solution found via the DeTurck method is necessarily a solution to the Einstein equation. On the other hand, we calculated the black hammocks in Bondi-Sachs gauge. As far as we know, this is the first use of such a gauge to find flowing AlAdS spacetimes in five dimensions. It seems as though Bondi-Sachs gauge works particularly well for spacetimes with a null hypersurface ``deep'' in the bulk, \textit{i.e.} opposite the conformal boundary in the integration domain. Such solutions, for example, the hammocks of this work, as well as flowing funnel solutions can also be found using the DeTurck method, though generally in these cases one cannot say with absolute certainty that the solutions are not Ricci solitons, and moreover, the solutions need to be found to a high degree of precision and on a large number of lattice points in order for quantities like the holographic stress tensor to be extracted. For example, we found that extracting the stress tensor of the black hammock solutions in DeTurck gauge required around twice the number of lattice points in the radial direction compared with when using the solutions found in Bondi-Sachs gauge. Not only this, but to find the hammocks in DeTurck gauge, one has to solve for seven metric functions, rather than just the five in Bondi-Sachs gauge. Moreover, in Bondi-Sachs gauge we were able to explore far more of the parameter space. It would be very interesting to explore the efficacy of the Bondi-Sachs gauge in this context in more detail, and we believe its use may open a door to solving other difficult problems, particularly those with the presence of a flowing horizon.

The horizon of the black hammock is not a Killing horizon, allowing it to have interesting properties, such as classical flow along it, as well as non-vanishing expansion and shear. This does not contradict the well-known rigidity theorems, since these apply only to geometries with horizons with compactly generated spatial cross-sections. The boundary event horizon is always hotter than the boundary cosmological horizon, and hence, as one would expect, we found that the classical flow along the black hammock is always from the event horizon to the cosmological horizon. The results of the shear and expansion are interesting, particularly the result that the shear vanishes at a point along the horizon, with all of its components swapping signs across this point.

It would be of great interest to be able to investigate the phase diagram of the CFT living on the de Sitter-Schwarzschild background. This would be achieved by deducing which of the gravitational duals dominates for a given value of the parameter, $\rho_h$. It would be expected that for very small $\rho_h$, when the boundary horizons are very far apart, the confined, tunnel phase would dominate, whilst for large $\rho_h \simeq 1$, where the horizons are almost aligned, the deconfined, hammock phase would dominate. Hence, the expectation is that there is a phase transition at some intermediate value of $\rho_h$. However, there is some difficulty in confirming this behaviour and finding the location of this phase transition.

In \cite{Santos:2020kmq}, it was suggested that a similar transition between black droplet and funnel solutions, which are dual to a CFT on a flat Schwarzschild background, occurs at the point in the parameter space at which the flow, $\Phi$, is maximised. We find a maximum of the flow, as seen in Figure \ref{fig:flow}, so one may conjecture that this is signifying the phase transition. However, the position of this maximum depends on the choice of quantity used to make the flow dimensionless. We used the temperature of the boundary event horizon, $T_H$. It is not clear to us that this is the meaningful quantity to use, as opposed to the temperature of the boundary cosmological horizon, $T_c$, or indeed some combination of these two temperatures.

This issue is linked to the difficulty to undergo a thermodynamic study of the two solutions in order to deduce which dominates. For example, in \cite{Marolf:2019wkz}, the phases of similar solutions was investigated by comparing the free energy, $F = E - TS$, of each of the solutions, with the solution with least free energy dominating. However, in our case, the solutions are not in thermal equilibrium which means that the free energy is very difficult to define. Put another way, it is not at all clear what value of the temperature, $T$, one should use to compute the free energy.

Moreover, we found there was some difficulty in extracting the difference in the entropy of the two solutions. In particular, we found that the difference between the sum of the areas of the two tunnel horizons and the area of the hammock horizon is infinite, even after matching the conformal frame. Divergences in extremal surface area can depend only upon the boundary metric, so this suggests that the hammock horizon is not the minimal extremal surface.\footnote{We would like to thank Don Marolf for his insight into this issue.} Therefore, in order to find the entropy difference, one would have to extend the geometry of the hammock through the bulk horizon, find the extremal surface and compare the area of it to the area of the two bulk horizons of the black tunnel. It would be very interesting to further understand this divergence of the difference between the horizon areas of the two solutions, and to explicitly find the extremal surface for the hammock.

In lieu of a thermodynamic argument to determine the dominance of either the tunnel or hammock, it appears that the best course to take in order to investigate the phase diagram may be to explicitly test the stability of the two bulk solutions across the parameter space. Fortunately, Bondi-Sachs gauge is extremely well-adapted to allow for one to slightly perturb the black hammock solutions and track over time-evolution whether this perturbation leads to an instability. One would perhaps expect that for small values of $\rho_h$, the black hammock would be unstable to such a perturbation, and hence would decay into a black tunnel. However, in order for such a transition to occur, the single bulk horizon of the hammock would have to ``pinch off'' and split into two disconnected horizons. Such evolution would violate the weak cosmic censorship conjecture \cite{Penrose:1969pc}, perhaps in a similar manner to the Gregory-Laflamme instability of the black string \cite{Gregory:1993vy, Lehner:2010pn}. It has previously been shown that weak cosmic censorship can be violated in AdS \cite{Niehoff:2015oga,Dias:2016eto,Horowitz:2016ezu,Crisford:2017zpi}, though the current set-up may prove to be an easier problem in which to explicitly find the evolution towards a naked singularity. We hope to address this issue in the near future.
\section*{Acknowledgements}
 We would like to thank \'Oscar~C.~Dias and Don~Marolf for reading an earlier version of this manuscript and for providing critical comments. J.~E.~S. has been partially supported by STFC consolidated grant ST/T000694/1 and W.~B. was supported by an STFC studentship, ST/S505298/1, and a Vice Chancellor's award from the University of Cambridge. The numerical component of this study was carried out using the computational facilities of the Fawcett High Performance Computing system at the Faculty of Mathematics, University of Cambridge, funded by STFC consolidated grants ST/P000681/1, ST/T000694/1 and ST/P000673/1.

\appendix

\section{Convergence properties for the black tunnels}\label{app:tunnels}
In section \ref{sec:tunnels}, we argued that due to the staticity of the black tunnels, Ricci solitons could not exist, \textit{i.e.} any solution to the Einstein-DeTurck equation would necessarily also be a solution to the Einstein equation. The DeTurck vector, $\xi^a$, therefore provides a good test of the convergence properties of the numerical method as one goes to the continuum limit.
\begin{figure}[b]
    \centering
    \includegraphics[width=0.75\linewidth]{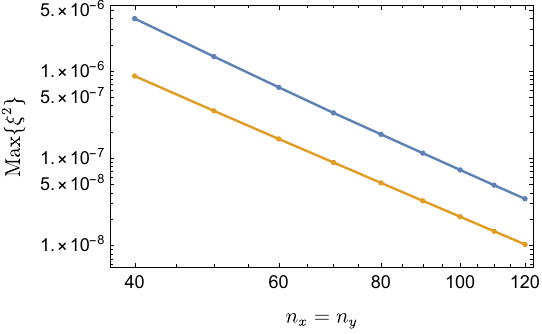}
    \caption{The maximum value of the norm of the DeTurck vector, $\xi^2=\xi^a\xi_a$, plotted in a log-log plot against the number of lattice points in each direction. The blue points correspond to the black tunnel solution with $\rho_h = 0.3$ and the orange points  correspond to the black tunnel solution with $\rho_h=0.9$. In each case $n_x=n_y$ where $n_i$ is the number of points used for the $i$-coordinate. The straight line in the log-log scale is indicative of power-law convergence.}
    \label{fig:drop_err}
\end{figure}
In Figure \ref{fig:drop_err}, we have plotted the maximum value of the norm of the DeTurck vector, given by $\xi^2 = \xi^a\xi_a$, across each of the lattice points. We have varied the number of lattice points used in the numerical method, keeping the ratio between the number of points in the $x$ and $y$ directions fixed at $1:1$. 

Clearly the norm of the DeTurck vector becomes smaller and smaller as the number of lattice points is increased, suggesting, as expected, that we are indeed approaching a solution to the Einstein equation. In Figure \ref{fig:drop_err} we have used a log-log scale in the axes to demonstrate that the convergence follows a power law behaviour. The reason the convergence is slower than exponential is because of the non-analytic behaviour arising in the solution, for example in the expansion near the conformal boundary described in (\ref{eq:drop_expansion}).

\section{Convergence properties for the black hammocks}\label{app:hammocks}
As discussed in section \ref{sec:BS}, in order to solve the Einstein equation for the hammock \textit{Ansatz}, we actually only solve a subset of the Einstein equation, specifically the bulk equations, $E_{ij}=0$, for $i,j\neq v$. We set the remaining components of the Einstein equation to zero, $E_{va}=0$, only as a boundary condition at $y=1$. If this is the case then the contracted Bianchi identity should ensure that in fact $E_{va}=0$ throughout the whole spacetime on a solution to the bulk equations. However, keeping track of $E_{va}$ throughout the bulk allows one to check the numerical method and its convergence properties as one increases the number of lattice points used in the discretization.

\begin{figure}[h!]
    \centering
    \includegraphics[width=0.75\linewidth]{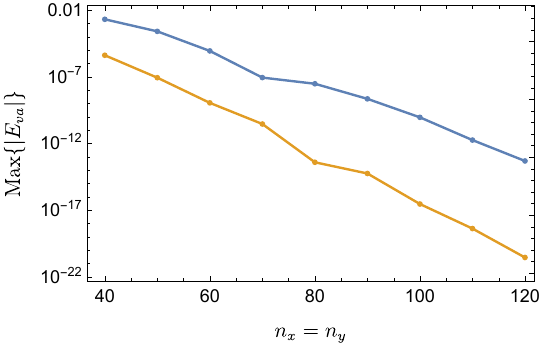}
    \caption{The maximum value of the components, $E_{va}$, of the Einstein equation that are not explicitly solved as bulk equations plotted in a log plot against the number of lattice points in each direction. The blue points  correspond to the black hammock solution with $\rho_h = 0.3$ and the orange points correspond to the black hammock solution with $\rho_h=0.9$. In each case $n_x=n_y$ where $n_i$ is the number of points used for the $i$-coordinate.}
    \label{fig:ham_err}
\end{figure}

In Figure \ref{fig:ham_err} we have plotted the maximum absolute value throughout the whole bulk of the $E_{va}$ components of the Einstein equation.
On the $x$-axis is the number of lattice points used in each direction in the discretization process, once again with the ratio of points in the $x$ and $y$ directions fixed at $1:1$.

This plot clearly shows that the maximum error in $E_{va}$ becomes very small as one increases the number of points, meaning that the numerical method converges very well to a solution to the Einstein equation. This time the plot is a log plot, that is, there is a log scale on the $y$-axis, but not the $x$-axis. Therefore, the fact that the plot seems to roughly give a straight line is indicative of exponential convergence, which is expected since no non-analytic terms arise in Bondi-Sachs gauge.

\vspace{200pt}
\bibliographystyle{JHEP}
\bibliography{papers}

\end{document}